\documentclass[a4paper,11pt]{article} 

\usepackage{jheppub} 

\usepackage[T1]{fontenc} 
\usepackage{physics}
\usepackage{physunits}
\usepackage{url}
\usepackage{mathtools}
\usepackage{gensymb}
\usepackage{multirow}
\usepackage{slashed}

\newcommand{\eg}{\textit{e.g.}}
\newcommand{\ie}{\textit{i.e.}}

\newcommand{\cL}{\mathcal{L}}

\newcommand{\cB}{\mathcal{B}}

\newcommand{\cO}{\mathcal{O}}

\newcommand{\cP}{\mathcal{P}}

\newcommand{\GeV}{\textrm{GeV}}

\newcommand{\beq}{\begin{equation}}
\newcommand{\eeq}{\end{equation}}

\usepackage[dvipsnames]{xcolor}

\preprint{MIT-CTP/5816}

\title{Searching for exotic scalars at fusion reactors}

\author[a]{Chaja Baruch,}
\author[b,c,a]{Patrick J. Fitzpatrick,}
\author[d]{Tony Menzo,}
\author[a]{Yotam Soreq,}
\author[e]{Sokratis Trifinopoulos,}
\author[d]{Jure Zupan}

\affiliation[a]{Physics Department, Technion--Israel Institute of Technology, Haifa 3200003, Israel}
\affiliation[b]{Dual CP Institute of High Energy Physics, C.P. 28045, Colima, M\'exico}
\affiliation[c]{Instituto de F\'{\i}sica, Universidad Nacional Aut\'onoma de M\'exico, A.P. 20-364, Ciudad de M\'exico 01000, M\'exico}
\affiliation[d]{Department of Physics, University of Cincinnati, Cincinnati, OH 45221, USA}
\affiliation[e]{Center for Theoretical Physics, Massachusetts Institute of Technology, Cambridge, MA 02139, USA}

\emailAdd{chajabaruch@campus.technion.ac.il}
\emailAdd{patrickf@campus.technion.ac.il}
\emailAdd{menzoad@mail.uc.edu}
\emailAdd{soreqy@physics.technion.ac.il}
\emailAdd{trifinos@mit.edu}
\emailAdd{zupanje@ucmail.uc.edu}

\abstract{
The energy created in deuterium-tritium fusion reactors originates from a high-intensity neutron flux interacting with the reactor's inner walls. 
The neutron flux can also be used to produce a self-sustaining reaction by lining the walls with lithium-rich `breeding blankets', in which a fraction of neutrons interacts with lithium, creating the tritium fuel. 
The high-intensity neutron flux can also result in the production of dark sector particles, feebly interacting light scalars or pseudoscalars, via nuclear transitions within the breeding blanket. 
We estimate the potential size of such dark sector flux outside the reactor, taking into account all current constraints, and consider possible detection methods at current and future thermonuclear fusion reactors. 
As a by-product, we also recast the SNO axion bound for a CP even scalar.
We find that year-long searches at current and future reactors can set leading constraints on dark scalar-- and dark pseudoscalar--nucleon couplings.
}

\begin{document} 
\maketitle
\flushbottom

\section{Introduction}
\label{sec:intro}

In recent years, the search for weakly coupled light particles produced via exotic nuclear transition emissions has emerged as a prominent avenue for the discovery of new physics~(NP).
Existing searches that have successfully placed stringent constraints on a number of NP scenarios have either focused on nuclear transitions that occur within the sun~\cite{Moriyama:1995bz,Krcmar:2001si,Derbin:2005xc,CAST:2009jdc,CAST:2009klq,Borexino:2012guz,Budnik:2019olh} or in nuclear fission reactors~\cite{Donnelly:1978ty,Zehnder:1981qn,Lehmann:1982bp,Cavaignac:1982ek,TEXONO:2006spf,Pospelov:2017kep,Benato:2018ijc,Dent:2019ueq,AristizabalSierra:2020rom,Massarczyk:2021dje,Arias-Aragon:2023ehh}. 
In this work, we discuss for the first time the possibility of producing light new physics in nuclear \textit{fusion} reactors.

Since the first demonstration of nuclear fusion in the 1950s, the design and development of nuclear fusion reactors has grown into an active field of research, for recent reviews, see, e.g., Refs.~\cite{meschini2023review,barbarino2020brief}.
Modern thermonuclear fusion reactors are designed primarily around the exothermic nuclear reaction 
\begin{align}
    \label{eq:DT_fusion}
    {^2_1\text{D}} + {^3_1\text{T}} 
    ~ \rightarrow ~ 
    {^4_2\text{He}} + n + 17.6\,\MeV \,,
\end{align}
where D, T, and He denote deuterium, tritium, and helium ($\alpha$-particle) nuclei, respectively, and $n$ is a neutron. 
The 17.6\,MeV energy excess is in the form of kinetic energy, split between the $\alpha$-particle ($\approx3.5\,\MeV$) and neutron ($\approx14.1\,\MeV$). 
The charged $\alpha$-particles remain confined and contribute to plasma heating, while neutrons stream freely to the reactor walls. 
The incident neutrons can transfer their energy to the reactor walls, generating heat that can ultimately be converted into electrical energy via conversion into the mechanical work generated by the reactor coolants. 
Modern reactors are designed with lithium-lined inner walls, referred to as the `breeding blankets', which interact with the incident neutrons to produce additional tritium nuclei that stream back into the main reactor volume, producing a self-sustained fusion reaction. 

\begin{figure}[t!]
    \centering
    \includegraphics[width=1.0\textwidth]{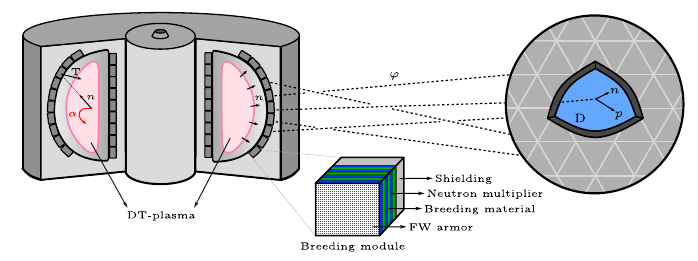}
    \caption{Schematic depiction of new physics production and detection in nuclear fusion facilities.}
    \label{fig:schematic_summary}
\end{figure}

The array of nuclear reactions occurring in the fusion reactors provide a number of possible mechanisms for the production of light new physics particles.  
If NP particles couple to nucleons, such interactions can lead to exotic particle production within the reactor volume as a correction to the fusion process itself, Eq.~\eqref{eq:DT_fusion}, in the form of initial and final state bremsstrahlung of NP particles. 
Alternatively, NP particles may be emitted from absorption or scattering of incident neutrons on nuclei in the reactor walls. 
That is, rather than interacting with the breeding blanket to produce additional tritium, neutron absorption can create an unstable nuclear state that quickly radiates to a stable configuration. 
If the NP particles couple to nucleons and/or photons, the decay of an unstable nuclear state may occur via an exotic nuclear transition, creating a NP particle in the final state. 
Both production mechanisms can create a detectable flux of NP particles just outside of the walls of the reactor, even if the NP particles are very weakly interacting. 
In the main part of the paper we estimate the NP flux that could arise from these production mechanisms, while being consistent with current constraints, and estimate the reach of possible detection methods. 

The paper is structured as follows. 
In Sect.~\ref{sec:benchmark} we introduce the NP models of interest. 
In Sect.~\ref{sec:production} we then consider individual production mechanisms and estimate the respective light new physics fluxes. 
In Sect.~\ref{sec:detection} we consider possible detection of light NP particles via deuterium dissociation, with the resulting sensitivity to light NP models estimated in Sect.~\ref{sec:results}. 
Conclusions in Sect.~\ref{sec:conclusion} are followed by several appendices containing additional details. 
App.~\ref{app:scalarD} contains details about the light scalar induced deuterium dissociation rates calculation. 
In App.~\ref{app:reactors} we collect short summaries of reactor parameters for a number of prominent fusion reactors that are either already in construction or are being proposed.  
App.~\ref{app:magnetic_conv} contains the details about possible searches using magnetic conversion, while App.~\ref{sec:toy:example} contains a toy example for scalar production via bremsstrahlung. 

\section{Benchmark new physics models}
\label{sec:benchmark}

To evaluate the sensitivity of fusion reactor based searches to light NP, we consider two benchmark NP models with light spin-0 particles: a light scalar and a light pseudo-scalar model. 

\paragraph{Light scalar.} 
The first example is a Standard Model~(SM) extended by a light scalar $\phi$. 
The corresponding interaction Lagrangian at low energies, $\mu\sim \cO(10\,\MeV)$, is given by\footnote{Note that we use the relativistic notation for nucleon currents even though the typical energy and momentum exchange in these experiments  is small, $|\vec q|\sim\cO(10\,\MeV)\ll m_N$. 
One could also have worked in the heavy-baryon formalism, in which the nucleon mass $m_N$ is integrated out~\cite{Jenkins:1990jv}. 
In that case the couplings to nucleons would be $g_{\phi N} \phi \, N_v^\dagger N_v$ and $g_{aN} a  \, N_v^\dagger q\cdot S_N N_v$, for Eqs.~\eqref{eq:Lphieff} and~\eqref{eq:Laeff}, respectively, where $S_\mu$ is the spin operator and $q_\mu$ the momentum exchange four-vector (for notation, see, e.g., Ref.~\cite{Bishara:2017pfq}).} 
\beq
    \begin{split}
    \label{eq:Lphieff}
    \cL_{\phi,{\rm eff}} 
    \supset  
    &-\frac{1}{4} g_{\phi\gamma} \phi F^{\mu\nu}F_{\mu\nu} +  
    \phi \sum_{N=p,n}  g_{\phi N} \bar N N+ g_{\phi e} \phi\, \bar e e\,,
\end{split}
\eeq
where $F_{\mu\nu}=\partial_\mu A_\nu -\partial_\nu A_\mu$ is the electromagnetic field strength tensor. 
The coupling $g_{\phi \gamma}$ is dimensionful, with units $\GeV^{-1}$, while $g_{\phi N}$ and $g_{\phi e}$ are dimensionless. 
The above effective Lagrangian relies on an implicit assumption that the couplings of $\phi$ to quarks are flavor diagonal. 
That is, we assumed that the couplings to quark and gluons are of the following form 
\beq
    \label{eq:Lphi:quarks}
    \cL_{\phi} 
    \supset 
    \sum_{q=u,d,s} g_{q}^\phi \, \phi\, \bar q q +\frac{\alpha_s}{12\pi}  g_{g}^\phi \phi G_{\mu\nu}^a G^{a\mu\nu}.
\eeq
The assumption of vanishing flavor violating couplings will be revisited below, when a UV model dependent comparison with the existing constraints from kaon decays will be made, see the discussion in Sect.~\ref{sec:results}.

The flavor-diagonal couplings to quarks and the couplings to the gluon field strengths in Eq.~\eqref{eq:Lphi:quarks} induce the couplings of $\phi$ to hadrons.  
The interactions that are relevant for the phenomenology of $\phi$ at the fusion reactors are the couplings of $\phi$ to neutrons, $n$, and protons, $p$, Eq.~\eqref{eq:Lphieff}.
Numerically, 
\begin{align}
    \begin{split}
    \label{eq:gphip:num}
    g_{\phi p}=& g_g^\phi F_G^p(0)+\sum_{q=u,d,s} g_q^\phi  \frac{F_S^{q/p}(0)}{m_q}
    \\
    =& -0.0504(6) \frac{g_g^\phi}{1\,\text{GeV}^{-1}}+7.5(1.2) g_u^\phi+6.5(9) g_d^\phi+0.46(5) g_s^\phi, 
    \end{split}
    \\
    \begin{split}
    g_{\phi n}=&g_g^\phi F_G^n(0)+\sum_q g_q^\phi  \frac{F_S^{q/n}(0)}{m_q}  
    \\
    =& -0.0504(6) \frac{g_g^\phi}{1\,\text{GeV}^{-1}}+6.7(1.0) g_u^\phi+7.3(9) g_d^\phi+0.46(5) g_s^\phi, 
    \end{split}
\end{align}
where we used the values for the form factors at renormalization scale $\mu=2\,\GeV$ from Ref.~\cite{Haxton:2024lyc} and the values of quark masses from Ref.~\cite{ParticleDataGroup:2024cfk}. 

Note that if all the couplings to the light SM fermions are of similar size, $g_u^\phi\sim g_d^\phi\sim g_s^\phi\sim g_{\phi e}$, then the induced couplings to nucleons are roughly an order of magnitude larger than the coupling to electrons, $g_{\phi N}\sim \cO(10) \times g_{\phi e}$.
In the UV complete light scalar models, though, the couplings of $\phi$ to the SM fermions are quite often proportional to the fermion masses, $g_\psi^\phi \propto m_\psi$ (an example is a light Higgs mixed scalar~\cite{Patt:2006fw,Beacham:2019nyx}). 
In that case the couplings to nucleons are even further enhanced,  $g_{\phi N}\sim \cO(10^2) \times g_{\phi e}$. 
Other hierarchies among flavor diagonal couplings are in general also possible, see, e.g., Refs.~\cite{Batell:2017kty,Batell:2018fqo,Batell:2021xsi,DiLuzio:2020oah,Balkin:2024qtf,Delaunay:2025lhl,DiLuzio:2023cuk}. 

\paragraph{Light pseudoscalar.} 
The second benchmark NP model is a SM extended by a light pseudo-scalar $a$, also referred to as the axion-like particle~(ALP), whose interactions are given by 
\beq
    \begin{split}
    \label{eq:Laeff}
    \cL_{a,{\rm eff}} \supset & 
    -\frac{1}{4} g_{a\gamma} a F^{\mu\nu}\tilde{F}_{\mu\nu} 
    + a \sum_{N=p,n} g_{aN} \big(\bar N i \gamma_5 N\big) + g_{a e} a\, \big(\bar e i \gamma_5 e\big)\,,
    \end{split}
\eeq
where $\tilde{F}_{\mu\nu} \equiv\frac{1}{2} \epsilon_{\mu\nu\alpha\beta}F^{\alpha\beta}$ is the dual of the electromagnetic field strength tensor, and we again assume flavor diagonal couplings to the SM fermions. 
Note that $g_{a\gamma}$ is a dimensionful coupling with units $\GeV^{-1}$, while $g_{aN}$ and $g_{ae}$ are dimensionless.
At the parton level, a commonly adopted notation for these couplings is (see, e.g., Ref.~\cite{Calibbi:2020jvd})
\beq
    \label{eq:L:a:quarks}
    \cL_{a}
    \supset 
   c_G^a \frac{a}{f_a}\frac{\alpha_s}{8\pi} G_{\mu\nu}^a \tilde G^{a\mu\nu}
    +\frac{\partial_\mu a}{2 f_a} \sum_{\psi=u,d,s,e} c_{\psi}^a \, \bar \psi \gamma^\mu \gamma_5 \psi\,,
\eeq
in terms of which the effective interactions in Eq.~\eqref{eq:Laeff} are given by
\beq
    g_{aN}
    =
    \frac{1}{f_a}\Big[\sum_{q=u,d,s} F_P^{q/N}(0) c_q^a-F_{\tilde G}^N(0) c_G^a\Big],\qquad 
    g_{ae}
    =
    \frac{m_e}{f_a} c_e^a\,,
\eeq
where for the nuclear matrix elements we used the notation of Refs.~\cite{Haxton:2024lyc,Bishara:2017pfq}. 
Numerically, $F_{\tilde G}^n(0)=F_{\tilde G}^p(0)=-0.40(5)\,\GeV$, $F_P^{u/p}(0)=F_P^{d/n}(0)=0.32(5)\,\GeV$, $F_P^{d/p}(0)=F_P^{u/n}(0)=-0.87(5)\,\GeV$, $F_P^{s/p}(0)=F_P^{s/n}(0)=-0.29(2)\,\GeV$, where we used the averages for the nuclear matrix elements from \cite{Haxton:2024lyc}. 
For anarchic couplings of $a$ to the SM fermions, $c_q^{a}\sim c_e^a\sim \cO(1)$, such as, e.g.,  in axiflavon models~\cite{Wilczek:1982rv,Davidson:1981zd,Davidson:1983fe,Calibbi:2016hwq,Ema:2016ops,Calibbi:2020jvd}, we therefore have $g_{aN}\sim \cO(\GeV/f_a) $ and  $g_{ae}\sim \cO(m_e /f_a)$, and thus $g_{ae}\sim 10^{-3} g_{aN}$. 

\paragraph{Decay modes.}
In the phenomenological analysis below we will mostly be interested in the mass ranges for the light spin-0 particles, $\varphi\equiv\phi,a$, where these are heavy enough so that they can decay to $e^+e^-$ pairs, but light enough so that their mass is below the $\mu^+\mu^-$ threshold. 
That is, we will work in the regime where only the $\varphi\to \gamma\gamma$ and $\varphi \to e^+e^-$ decay channels are open.

The decay widths for the two channels are ($m_{\varphi} > 2 m_e$)
\begin{equation}
    \Gamma(\varphi \to \gamma \gamma) 
    = 
    \frac{g^2_{\varphi\gamma, {\rm eff}} m_{\varphi}^3}{64\pi}, 
    \quad \quad 
    \Gamma(\varphi \to ee) 
    = 
    \frac{g^2_{\varphi e} m_{\varphi}}{8\pi} 
    \sqrt{1 - \frac{4m_e^2}{m_\varphi^2}} \, , 
\end{equation}
where\footnote{Note that the couplings to photons and electrons are defined at $\mu\sim 10\,\MeV$, with neutron and proton fields in Eqs.~\eqref{eq:Lphieff} and~\eqref{eq:Laeff} implicitly treated as heavy. The contributions from pion loops and heavier hadronic states are thus already included in $g_{\varphi \gamma}$, along with any nonzero UV contributions.}
\beq
    \label{eq:g:varphi:gamma}
    g_{a\gamma,{\rm eff}} 
    = 
    g_{a\gamma}- \frac{\alpha}{\pi} \frac{g_{ae}}{m_e} B_1(1/\tau_e)\, , 
    \qquad 
    g_{\phi\gamma,{\rm eff}} 
    =  
   g_{\phi\gamma}- \frac{2}{3}\frac{\alpha}{\pi} \frac{g_{ae}}{m_e} A_f(\tau_e)\,.
\eeq
where $\tau_e=m_\phi^2/4 m_e^2$. 
The pseudoscalar loop function is $B_1\simeq 1$ for heavy $a$, $m_a\gg m_e$, and is $B_1\simeq - m_a^2/(12 m_e^2)$ for light $a$, $m_a\ll m_e$, see, e.g., Ref.~\cite{Bauer:2020jbp}. 
The scalar loop function $A_f(\tau_e)$ can be found, e.g., in Ref.~\cite{Carmi:2012in} and is $A_f(\tau_e)\simeq 1$ for $\tau_e\ll 1$, i.e., for $2m_e\gg m_\phi$.

For $\varphi$ masses above the $e^+e^-$ threshold we generically expect that $\Gamma(\varphi \to e^+e^-) \gg \Gamma(\varphi \to \gamma\gamma)$, since couplings to photons are loop suppressed. 
Therefore, the dominant decay mode of $\varphi$ is expected to be $\varphi\to e^+e^-$, if this is kinematically allowed. 
The decay time is given by (not displaying the phase space factor for clarity) 
\beq
    \label{eq:ctau:electrons}
    c \tau_\varphi
    \simeq 
    5.0 \,\text{km} \left(\frac{10^{-8}}{g_{\varphi e}}\right)^2
    \left(\frac{10\,\MeV}{m_\varphi}\right)\,,
\eeq
where in the numerical example we used $g_{\varphi e}=10^{-8} $ as an illustrative benchmark (as we will see in Sect.~\ref{sec:results} a typical sensitivity of light NP searches at future fusion reactors is expected to be $g_{ap}\sim \cO(10^{-5})$ and $g_{\phi p}\sim \cO(10^{-7})$, with the couplings to electrons, $g_{\varphi e}$,  a few orders of magnitude further suppressed, as discussed above). 
This means that unless $g_{\varphi e}$ is significantly larger than the naive expectations, $\varphi$ will not decay inside the reactor volume and will be able to reach a nearby detector.

The same is true for $\varphi$ masses below the $e^+e^-$ threshold, where the dominant decay channel is $\varphi \to \gamma\gamma$, with the corresponding lifetime given by 
\beq
    c \tau_\varphi
    \simeq 
    4.0 \times 10^{12}\, \text{m} 
    \left(\frac{10^{-7}\,\GeV^{-1}}{g_{\varphi\gamma,{\rm eff}}}\right)^2
    \left(\frac{0.1\,\MeV}{m_\varphi}\right)^3,
\eeq
and thus $\varphi$ also does not decay within the reactor volume. 

\paragraph{Numerical benchmarks.} 
For the fusion reactor phenomenology of light spin-0 particles considered in this work, the most important couplings are those to nucleons, $g_{\varphi N}$. 
The exact values of couplings to electrons and photons, $g_{\varphi e}$ and $g_{\varphi \gamma}$, are less important, as long as they are small enough so that the spin-0 particles do not decay inside the reactor volume, see the estimates above. 
However, they are important for other existing constraints. 
We thus consider two one-parameter benchmarks 
\begin{align}
    \label{eq:benchmark:leptophobic}
    \text{$p$-only\,:}
    &\qquad& 
    g_{\varphi p}\ne 0, \qquad g_{\varphi e}&=g_{\varphi \gamma,\text{eff}}=g_{\varphi n}=0, \\
    \label{eq:benchmark:electrons}
    \text{$pe\gamma$\,:}
    &\qquad &  
    g_{\varphi p}\ne 0, \qquad g_{\varphi e}&=10^{-3} g_{\varphi p}, \qquad  g_{\varphi \gamma,\text{eff}}= \frac{10^{-3}}{\text{GeV}}g_{\varphi p}, \qquad  g_{\varphi n}=0,
\end{align}
that we use in the numerical analysis below. 
In ``$p$-only'' benchmark we set all the couplings to zero, except couplings to protons. 
This is a minimal set-up for our fusion-reactor-based signal. 
In the ``$pe\gamma$'' benchmark, we also keep couplings to electrons and photons nonzero. 
The coupling to electrons is set to roughly the value $(m_e/m_p)\times g_{ap}$, and the couplings to photons to roughly $(\alpha/4\pi)\times (1/m_e)\times g_{ae}$, as motivated by UV origins of the two couplings, following the discussion above. 
To further simplify the discussion, we will assume that the couplings of $\varphi$ to neutrons are suppressed, which then also avoids  the constraints from neutron-nucleus scattering experiments~\cite{Barbieri:1975xy,Nesvizhevsky:2007by,Kamiya:2015eva,Haddock:2017wav,Heacock:2021btd}. 

\section{Production of light particles}
\label{sec:production}

\begin{figure}[t!]
    \centering
    \includegraphics[width=0.9\textwidth]{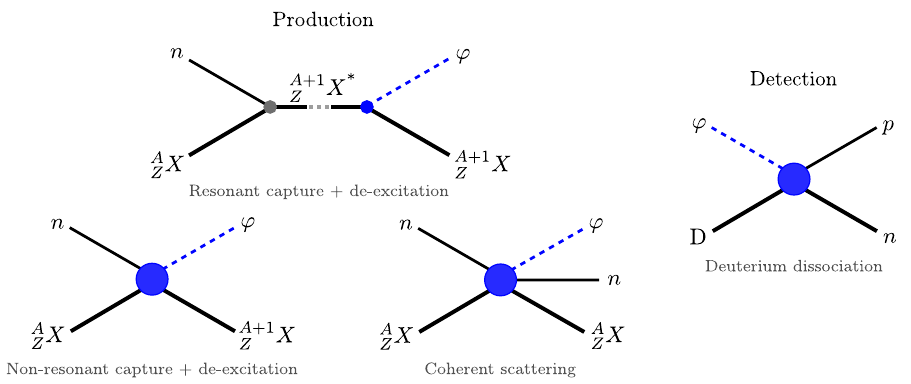}
    \caption{Spin-0 particle production (left) and detection (right) mechanisms.}
    \label{fig:production}
\end{figure}

\subsection{Production of light exotic spin-0 particles}

In this work, we focus on neutron absorption and neutron scattering as the main production mechanisms of exotic light spin-0 particles, as seen in Fig.~\ref{fig:production} (left). 
The exotic neutron absorption can be resonant
\begin{equation}
    \label{eq:neutron:abs:res}
     n + {^{A}_{Z}X} 
     \rightarrow {^{A+1}_{Z}X^*} \rightarrow 
     {^{A+1}_{Z}X} + \varphi\,,
\end{equation}
or non-resonant 
\begin{equation}
     n + {^{A}_{Z}X} \rightarrow  {^{A+1}_{Z}X} + \varphi\,,
\end{equation}
where ${^{A}_{Z}X}$ denotes an element $X$ of mass number $A$ and atomic number $Z$. 
The energy of the emitted $\varphi$ is given by
\beq
    E_\varphi=E_n +S_n,
\eeq
where $E_n$ is the energy of the incident neutron, while $S_n$ is the neutron separation energy for state ${^{A+1}_{Z}X}$ (the amount of energy required to strip ${^{A+1}_{Z}X}$ of a single neutron). 
The benefit of neutron capture is that the energy of the produced light exotic spin-0 particles can be significant, $E_\varphi\sim \cO(10\,\MeV)$, see Table~\ref{tab:res_n_capture}, even if the incident neutron energy is very small, below an MeV. 

We also estimate the production of $\varphi$ particles from neutron scattering 
\begin{equation}
    \label{eq:neutron:scatter}
     n + {^{A}_{Z}X} 
     \rightarrow  {^{A}_{Z}X} + n+\varphi\,.
\end{equation}
While this is a sub-leading production mechanism, it is still an important one for the case of a light scalar $\phi$. 
Furthermore, our estimates for light scalar $\phi$ production rates via neutron absorption carry very large uncertainties, while the production of $\phi$ in neutron scattering is theoretically under better control.
The downside is that only part of the incident neutron flux can be utilized, with energies $E_n$ above $2.2\,\MeV$, in order to generate an observable signal in deuteron dissociation. 

\begin{figure}
    \centering
    \includegraphics[width=0.65\textwidth]{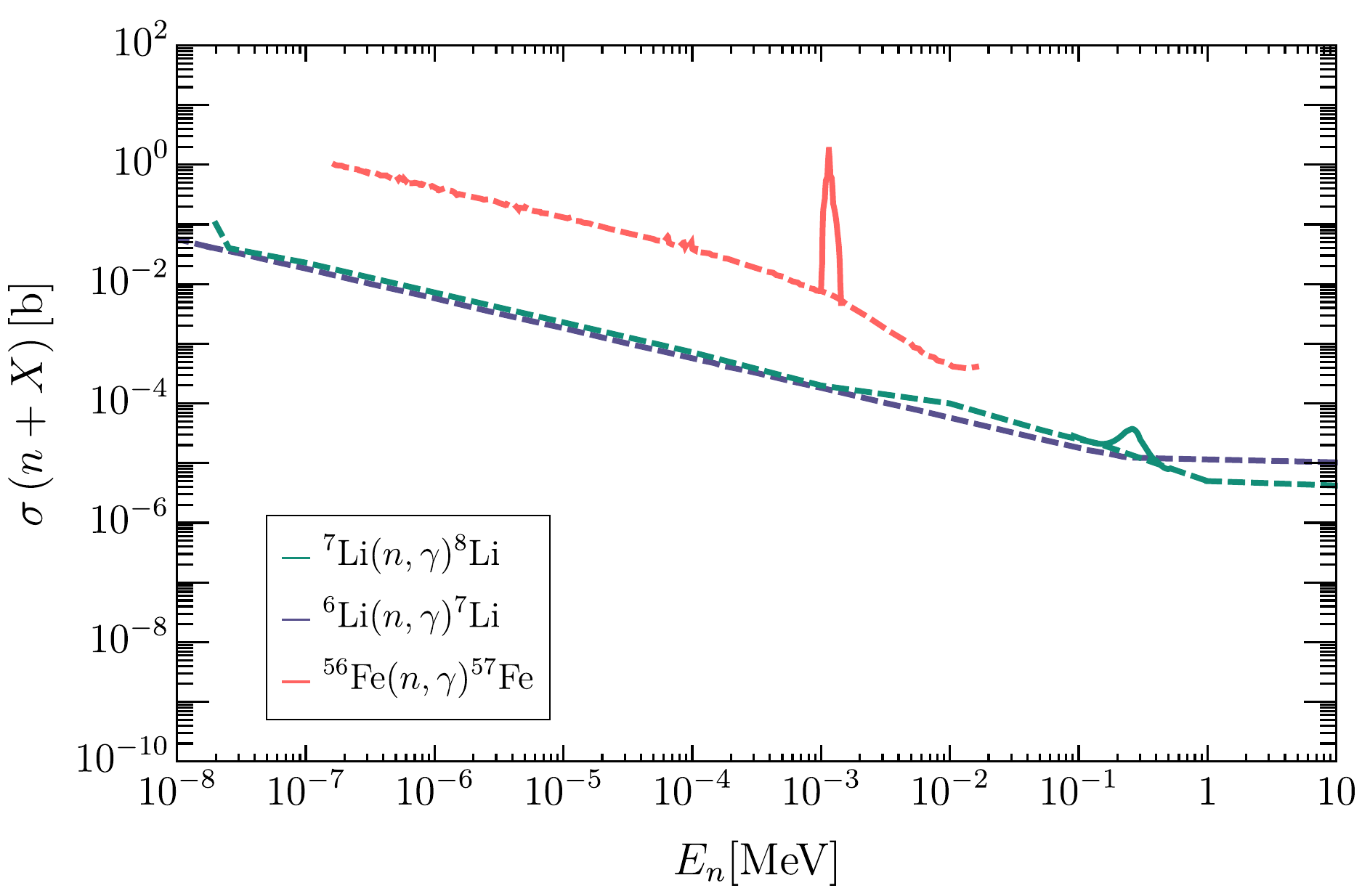}
    \caption{The $^7$Li, $^6$Li and $^{56}$Fe neutron capture cross sections extracted from Ref.~\cite{kopecky1997atlas}, with E1\,(resonant M1) transition dominated rates denoted with  dashed\,(solid) lines. See Eq.~\eqref{eq:M1:nonres} and the accompanying discussion regarding non-resonant M1 transitions. For $^7$Li the analysis regarding the separation of the E1, M1 contributions is taken from Refs.~\cite{heil1998n,fernando2012leading,lynn1991direct}.
    For the Fe cross section we assume that the resonance at $E_n=14.4\keV$ is due to an M1 transition. 
    We have no clear M1 resonance for ${}^6$Li.}
    \label{fig:capture_xsec}
\end{figure}

\begin{figure}
    \centering
    \includegraphics[width=0.65\textwidth]{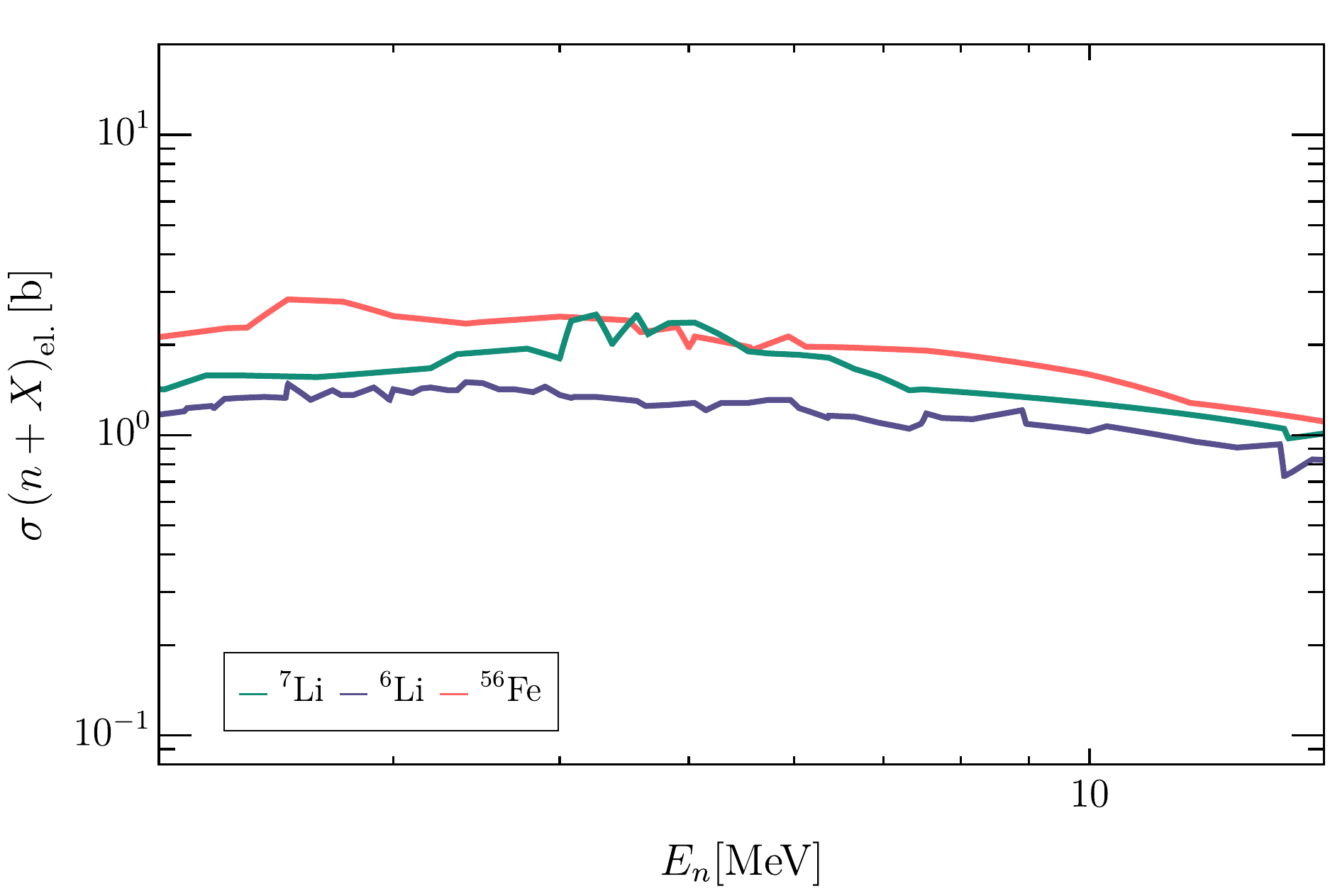}
    \caption{The cross sections for neutron scattering on $^7$Li, $^6$Li and $^{56}$Fe, taken from the EXFOR database~\cite{OTUKA2014272}.
    The neutron energy  range shown is the one relevant for detection via deuterium dissociation, \ie, $E_n\geq 2.2 \eV[M]$.}
    \label{fig:capture_xsec_elastic}
\end{figure}

Next, we give estimates for the corresponding cross sections.

\paragraph{Pseudoscalars.}
The resonant neutron absorption production of light pseudo-scalars can be related to the respective magnetic nuclear de-excitation rate using multipole techniques, working in the non-relativistic limit for the nucleon currents~\cite{Donnelly:1978tz}.\footnote{Here, the non-relativistic reduction refers to the expansion of the one-body nuclear matrix elements to leading order in $|p_N|/m_N \simeq v_N/c$ where $p_N, m_N,$ and $v_N$ are, respectively, the momentum, mass, and velocity of the nucleon that is participating in the decay of the excited nuclear state.}
For M1 processes\footnote{We use the standard notation, where the radiative electric E$J$ and magnetic M$J$ transitions are sourced by the transverse electric and magnetic nuclear response operators of $J-$th multipole order (dipole ($J = 1$), quadrupole ($J=2$), etc.). Conservation of parity requires E$J$ transitions to produce emissions with parity $P = (-1)^J$ and likewise M$J$ transitions to produce emission with parity $P = (-1)^{J+1}$. Similarly, the Coulomb multipoles are denoted as C$J$, see, e.g., Ref.~\cite{Donnelly:1975ze} for further details.} the ratio of the decay rates for de-excitation via pseudoscalar emission ($\Gamma_a$), and the SM photon emission ($\Gamma_\gamma)$ can be written as~\cite{Donnelly:1978ty,Avignone:1988bv}
\begin{align}
    \label{eq:M1_flux}
    \mathcal{B}_{a/\gamma;\text{M1}} \equiv \left(\frac{\Gamma_a}{\Gamma_\gamma} \right)_{\rm M1}
    =  
    \left(\frac{p_a}{E_\gamma}\right)^3 \frac{1}{2\pi\alpha} \frac{1}{1+\delta^2} \left(\frac{\left(1+\beta\right)g_{ap}}{\left(\mu_0-\frac{1}{2}\right)\beta + \mu_3 - \eta}\right)^2\,,
\end{align}
where we set the pseudoscalar-neutron coupling to zero, $g_{an}=0$, while $\mu_0\simeq 0.88$ and $\mu_3\simeq 4.71$ are the isoscalar and isovector magnetic moments in terms of nuclear magnetons, while $p_a=(E_a^2-m_a^2)^{1/2}$ is the outgoing ALP three-momentum, and $E_\gamma(=E_a)$ the photon\,(ALP) energy. 
The parameters $\eta$, $\beta$ and $\delta$ encode the differences in the responses of the nucleus to the pseudoscalar vs. electromagnetic currents; $\eta$ and $\beta$ are the nuclear structure parameters, while $\delta$ is the E2/M1 mixing ratio. 
Their values for the relevant nuclear transitions are listed in Table~\ref{tab:nuc_param}.

\begin{table}[t!]
    \centering
    \begin{tabular}{cccc}
        \hline 
        \hline
          & $^7$Li~\cite{CAST:2009klq} & $^8$Li~\cite{Waites:2022tov} & $^{57}$Fe~\cite{Massarczyk:2021dje} \\
          \hline
         $\eta$ & $0.5$ & $-0.1034$ & $0.8$ \\
         $\beta$ & $1.0$ & $1.0$ & $-1.19$\\
         $\delta$ & $0$ & $0$ & $0.002$ \\
         \hline 
         \hline
    \end{tabular}
    \caption{Nuclear structure parameters for the radiative M1 transitions, ${}^{A+1}_ZX\to {}^{A}_ZX+\gamma$, used in Eq.~\eqref{eq:sigma:pseudo} to estimate the resonant neutron absorption induced ALP emission rates, Eq.~\eqref{eq:neutron:abs:res}.}
    \label{tab:nuc_param}
\end{table}

The cross section for production of light pseudo-scalars from resonant neutron absorption is thus given by
\beq
    \label{eq:sigma:pseudo}
    \sigma_{a;\text{M1}}
    =
    \cB_{a/\gamma;\text{M1}}  \sigma_{\gamma;\text{M1}},
\eeq
where $\sigma_{\gamma;\text{M1}}$ is the cross section for the M1 production of $\gamma$'s, i.e., for the process $n+ {^{A}_{Z}X} \to {^{A+1}_{Z}X} + a$. 
These are shown with solid lines in Fig.~\ref{fig:capture_xsec} for neutron absorption on $^7$Li, $^6$Li and $^{56}$Fe, taken from NGATLAS~\cite{kopecky1997atlas}, and are then used in Eqs.~\eqref{eq:sigma:pseudo} and \eqref{eq:M1_flux} to obtain the pseudoscalar production rates. 

More precisely, neutron absorption on ${}^7$Li, followed by photon emission, is a sum of a non-resonant E1 (dashed green line) and a resonant M1 transition (solid green line)~\cite{heil1998n,fernando2012leading,lynn1991direct}, and similarly for neutron absorption on ${}^{56}$Fe (red lines). 
For $^{56}$Fe there is a very clear resonance at $E_n=14.4\,\keV$ which yields $^{57}$Fe$^\ast$ that de-excites purely through an M1 transition. 
Since there are no other clear resonances, we assume that the rest of the spectrum is entirely due to E1 transitions.
For light scalar production via neutron absorption on $^6$Li the situation is less clear. 
In Ref.~\cite{dong2017gamow}, using a Gamov shell model, it was predicted that there should be a relatively sizable M1 resonance comparable to the experimental errors but it is not yet clearly visible in data. 
Additionally, Ref.~\cite{C:2022uvr} finds a highly suppressed non-resonant M1 rate from a calculation based on  {\sc FRESCO} coupled channel code. 
In summary, we have no clear resonant M1 transitions for ${}^6$Li.

Note that using only the clearly visible M1 resonances to estimate the light pseudo-scalar production via neutron absorption is conservative. 
For instance, a naive order of magnitude estimate for the non-resonant M1 transition is 
\beq
    \label{eq:M1:nonres}
    \sigma_{\gamma;\text{M1}}^{\rm non-res.}
    \sim 
    (Q/m_N)^2 \sigma_{\gamma;\text{E1}}^{\rm non-res.},
\eeq
with $Q\sim 250\,\MeV$ a typical inverse nuclear size (i.e., the typical nuclear momentum, comparable to Fermi momentum), and thus $(Q/m_N)\approx 1/4$~\cite{Donnelly:1978tz}. 
If the non-resonant M1 neutron absorption rates follow this naive estimate, the pseudo-scalar rate would be $\cO(10^2)$ times larger than given using our nominal conservative approach, where we neglect the non-resonant contributions. 
In the numerical analysis we estimate the effect of such possible enhancement of the ALP flux, though those results should be viewed as only indicative, until confirmed by dedicated nuclear physics calculations.

\paragraph{Scalars.}
The estimates of light scalar production rates are significantly more uncertain. 
In contrast to light pseudo-scalar production which, in the long wavelength limit, can be related to the combination of M1 and E2 photon emissions, Eq.~\eqref{eq:M1_flux}, we do not have such a simple rescaling rule relating the emission of a light scalar $\phi$ to electromagnetic transitions. 
This is, to some extent, a problem of measurements. 
The C$J$ Coulomb emissions of an $e^+e^-$ pair in nuclear reactions could be used to obtain the emission rates of light scalars $\phi$. 
Such measurements are not available, though.
Without detailed nuclear calculations we use naive dimensional analysis~(NDA) to estimate the typical expected sizes of nuclear matrix elements of scalar currents. 

For $\phi$ production via neutron absorption we use, as in the ALP case, a comparison with the E1 electromagnetic emission rates. 
However, this comparison now faces several additional sources of uncertainty.  
The main challenge is that the unsuppressed C0 scalar emissions can only arise in elastic nuclear transitions, while inelastic emissions only start at C2 order. 
That is, the relatively enhanced C0 emissions of a light scalar can occur either from nuclei on the external legs in the Feynman diagram, while scalar emissions via de-excitations of the intermediate excited nuclear states to the ground state are highly suppressed.  
This is quite different from the electromagnetic emissions, where both types of contributions can in principle be part of the E1 emissions. 
Using the NDA estimates for the typical relative sizes of the C0 and E1 transitions from Table~3.1 in~\cite{Donnelly:1978tz} then gives
\beq
    \label{eq:scalar:n:abs}
    \sigma({{}^A_Z X(n,\phi){}^{A+1}_Z X})
    \approx 
    \sigma({{}^A_Z X(n,\gamma){}^{A+1}_Z X})\big|_{\text{E1}}^{\rm non-res.} 
    \left(\frac{g_{\phi p}}{e}\right)^2 
    \left(\frac{Q}{E_\gamma}\right)^2 
    \frac{p_\phi}{E_\gamma},
\eeq
where as in Eq.~\eqref{eq:M1:nonres} $Q\sim 250\,\MeV$ is the inverse of a typical nuclear radius, and the last factor in Eq.~\eqref{eq:scalar:n:abs} is due to the reduced phase space for massive $\phi$ with $p_\phi=(E_\phi^2 -m_\phi^2)^{1/2}$ the three-momentum of the outgoing scalar, and $E_\gamma(=E_\phi)$  the photon (scalar) energy, see App.~\ref{sec:toy:example}. 
As for the case of the ALP, we also assumed that $\phi$ only couples to protons, and thus set $\phi$ coupling to neutrons to zero, $g_{\phi n}=0$.

The above NDA estimate is expected to suffer from large uncertainties. 
In particular, we should not expect that the scalar emissions follow the ${}^A_Z X(n,\gamma){}^{A+1}_Z X$ E1 resonance structure.  
In our case, this is not an issue since all E1 emissions are non-resonant (dashed lines in Fig.~\ref{fig:capture_xsec_elastic}). 
Still, in the absence of detailed nuclear calculations for light scalar emissions in neutron absorption processes, the above estimate in Eq.~\eqref{eq:scalar:n:abs} should be taken only as a rough guidance. 

The light scalars can also be produced in neutron scattering, Eq.~\eqref{eq:neutron:scatter}. 
For the scattering rate we use an NDA estimate 
\beq
    \label{eq:sigm:nX}
    \sigma({{}^A_ZXn\to {}^{A}_ZX}n\phi)
    \approx
    \sigma({{}^A_ZXn\to {}^{A}_ZX}n)\big(Z g_{\phi p}\big)^2 \frac{}{} \frac{(\bar E_\phi^2-m_\phi^2)^{3/2}}{16\pi^2 \bar E_\phi^3},
\eeq
where as before we assumed that the coupling of $\phi$ to neutrons is negligible. For ${}^A_Z X={}^6\text{Li},{}^7\text{Li},{}^{56}\text{Fe}$, the cross sections $\sigma({{}^A_ZX n\to {}^{A}_Z X}n)$, which enter \eqref{eq:sigm:nX}, are shown in Fig.~\ref{fig:capture_xsec_elastic}.
The last factor in Eq.~\eqref{eq:sigm:nX} gives a naive estimate of the phase space suppression factor in the soft scalar limit, see App. \ref{sec:toy:example}.
Here, $\bar E_\phi$ is the typical energy of the outgoing scalar. 
As a rough estimate, we use $\bar E_\phi \approx E_n$, with $E_n$ the energy of the incoming neutron. 
While the average $E_\phi$ needs to be smaller than $E_n$ by definition, setting $\bar E_\phi =E_n$ in numerical estimates has only a minor impact on the magnitude of the phase space factor, while ensuring the correct value for the maximum $m_\phi$ that can be produced in the scattering.

\begin{table}[t]
\centering
\renewcommand{\arraystretch}{1.5}
\begin{tabular}{@{}ccccc@{}}
\hline\hline
\textbf{Process} & $\boldsymbol{S_n} ({^{A+1}_Z X}) $ & \textbf{Transition}   & \textbf{Multipolarity}                             & $\boldsymbol{E_\gamma}$ \\ \hline
\multirow{2}{*}{\({}^6\text{Li}(n,\gamma){}^7\text{Li}\)} & \multirow{2}{*}{7.25 MeV } & $|1/2^+\rangle \to |3/2^-\rangle$ & E1 (non-res.)   & 7.25 MeV  \\
                           &                       & $|1/2^+\rangle \to |1/2^-\rangle$ & E1 (non-res.) & 6.77 MeV      
                           \\ 
                           \hline
\multirow{2}{*}{\({}^7\text{Li}(n,\gamma){}^8\text{Li}\)} & \multirow{2}{*}{2.03 MeV

} & $|3/2^-\rangle \to |2^+\rangle$ & E1 (non-res.)   & 2.03 MeV    \\
                           &                       & $|3/2^-\rangle \to |3^+\rangle$ & M1 (res.) & 2.26 MeV    
                           \\ 
                           \hline
\multirow{4}{*}{\({}^{56}\text{Fe}(n,\gamma){}^{57}\text{Fe}\)} & \multirow{4}{*}{7.65 MeV
}  & $ |1/2^+\rangle \to |1/2^-\rangle$ & E1 (non-res.)  & 7.65 MeV \\ 
                           &                       & $|1/2^+\rangle \to |3/2^-\rangle$ & E1 (non-res.)   & 7.63 MeV \\ 
                           &                       & $|1/2^-\rangle \to |1/2^-\rangle$ & M1 (res.)  & 7.64 MeV \\
                           &                       & $|1/2^-\rangle \to |3/2^-\rangle$ & M1 (res.)  & 7.63 MeV \\ 
\hline\hline
\end{tabular}
    \caption{Summary of neutron capture processes we consider. The second column lists the neutron separation energy, the third column the $J^P$ quantum numbers of the initial and final states (if the transition is resonant, the quantum numbers are those of the intermediate resonance, for the nonresonant case they are those of the $L=0$ initial state). The fourth column lists multipolarity of the transition, and the final column gives the energy of the photon in the radiative transition.}
    \label{tab:res_n_capture}
\end{table}

\begin{figure}
    \centering
    \includegraphics[width=0.65\textwidth]{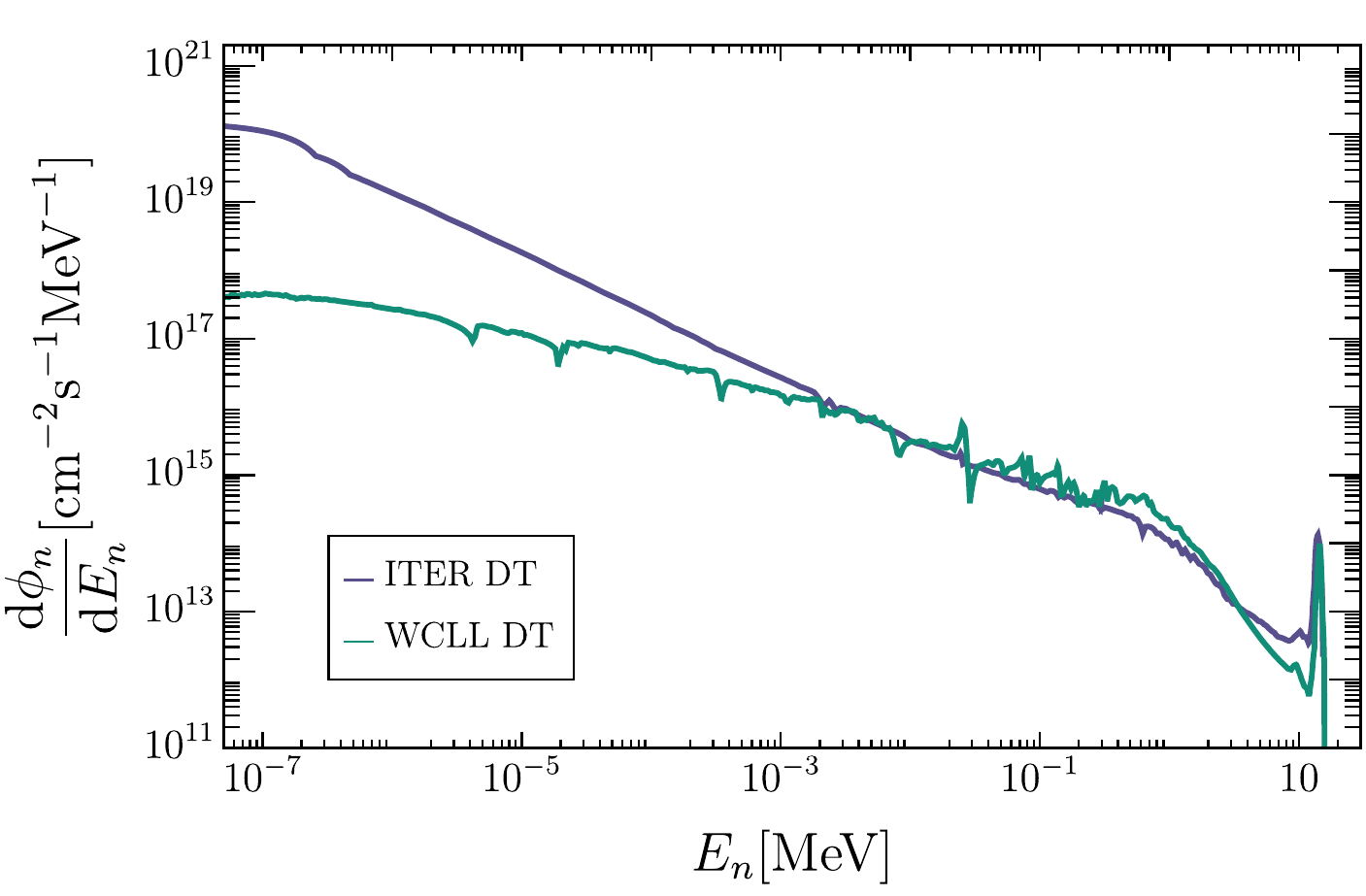}
    \caption{The incident neutron flux per energy  at the first wall of the reactor as a function of  the incident neutron energy, $E_n$. We show the expected neutron flux for two sample cases, taken from Ref.~\cite{fleming2018fispact}: ITER DT and WCLL DT, with their respective power outputs normalized to $P\simeq 2000$ MW. In the numerical analysis, we use ITER DT as the benchmark value.  } 
    \label{fig:neutron_flux}
\end{figure}

\subsection{Fusion reactors}
\label{subsec:reactor_summary}

In this work, we consider tokamak-based, magnetically confining fusion reactors utilizing deuterium-tritium fuel. 
These reactors feature a hollow vacuum vessel that maintains the low-pressure environment required for plasma confinement. 
Beyond the vacuum vessel, modern fusion reactors are typically surrounded by multiple specialized containment walls designed to withstand the extreme conditions of fusion while simultaneously allowing for efficient fuel production and energy transfer. 
The plasma facing first walls~(FW) typically feature a reactor-dependent thin layer of ``armor'' possessing low sputtering rates under plasma interaction and engineered to withstand high neutron fluences and extreme thermal loads. 
Beyond the FW, structural support layers, typically composed of reduced-activation ferritic-martensitic steels, interlaced with coolant channels, provide mechanical support and thermal management. 
Surrounding and embedded within these structural layers are breeding blanket modules that house alternating layers of neutron-multiplying materials, such as beryllium or lead, and tritium-breeding materials such as the lithium isotopes ($^6$Li and $^7$Li), which react with neutrons to sustain the fusion fuel cycle. 
In some designs, the system is terminated by a thick shield or reflector plate, redirecting escaping neutrons back toward the breeding layers to maximize fuel production. 
For a schematic diagram of a modern fusion reactor, see Fig.~\ref{fig:schematic_summary}.

The specific design and materials of these components vary significantly between reactors, tailored to meet each reactor's objectives and reflecting the ongoing nature of research and development within the field. 
For our analysis we consider reactor parameters comparable to that of future demonstrator~(DEMO) fusion reactors, the next generation offspring of the International Thermonuclear Experimental Reactor~(ITER) currently under construction in France and expected to be fully operational in the mid to late 2030s.
While ITER could, in principle, be sensitive to new physics, the bounds derived here are primarily relevant for reactors that will feature fully operational breeding blankets and can be designed with sufficient space to accommodate large-scale detectors for new physics experiments--criteria that ITER does not meet, as it will only have a partial blanket and limited space for auxiliary detectors.
These reactors expect to generate plasmas with thermal power outputs of $P = 500-2000\,$MW with an average expected neutron flux at the FW of
\begin{equation}
    \Phi^{\text{total}}_n 
    \simeq 
    \frac{P}{14.1\,\MeV} 
    \approx 10^{15} \text{ neutrons }\text{cm}^{-2} \text{s}^{-1} 
    \left( \frac{P}{2000 \text{ MW}} \right)\,.
\end{equation}
For further details about the current and future fusion reactor projects, including the discussion of design features and material choices that influence new physics production rates, see App.~\ref{app:reactors}. 

\subsection{Estimating the new physics flux}

The flux of exotic spin-zero particles at a detector placed a distance $L$ away from the first wall of the reactor is given by
\begin{equation}
    \label{eq:NP_flux}
    \frac{\dd\Phi_{\varphi}}{\dd E_n} 
    = 
    \frac{1}{4\pi L^2} \times \cP_{\text{surv.}} \times \frac{\dd\dot{N}_{\varphi}}{\dd E_n}\,,
\end{equation}
where for simplicity we treated the reactor as a point source, and $\dot{N}_{\varphi}$ is the total $\varphi=a,\phi$ production rate from the reactor. 
The survival probability, encoding the probability of $\varphi$ not to decay before it reaches the detector, is given by $\cP_{\text{surv.}} = e^{-m_{\varphi} L/|\vec{p}_{\varphi}|\tau_{\varphi}}$, where $\tau_\varphi$ is the $\varphi$ lifetime (see Sect.~\ref{sec:benchmark}) and $|\vec p_\varphi|$ its momentum. For the parameter space we consider $\cP_{\text{surv.}} \simeq 1$. 

The NP production rate per neutron energy is given by
\begin{equation}
    \label{eq:NP_num}
    \frac{\dd\dot{N}_{\varphi}}{\dd E_n} 
    = 
    \sum_{i,X} N_{X} \dv{\Phi_n}{E_n}~\sigma_\varphi^{i,X}(E_n) \,,
\end{equation}
where the sum runs over all the contributing production channels ($i$) and nuclei ($X$), $N_X$ is the effective number of atoms contained in the wall of the reactor that are accessible for interaction with the incoming neutrons, $\Phi_n$ is the neutron flux at the first wall of the reactor, and $\sigma^{i,X}_{\varphi}(E_n)$ the relevant cross-section for production of $\varphi$ particles; either $\sigma_{a;\text{M1}}$ or $\sigma_{\gamma;\text{M1}}^{\rm non-res.}$ in Eqs.~\eqref{eq:sigma:pseudo}, \eqref{eq:M1:nonres} for ALP production, or
$\sigma({{}^A_ZX(n,\phi){}^{A+1}_ZX})$ and $\sigma({{}^A_ZXn\to {}^{A}_ZX}n\phi)$ in Eqs.~\eqref{eq:scalar:n:abs}, \eqref{eq:sigm:nX} for scalar production. 

A precise estimate of $\dd\dot{N}_\varphi/\dd E_n$ requires a comprehensive neutronics analysis incorporating the reactor geometry, materials, and three-dimensional neutron spectrum embedded into a sophisticated neutron transport simulation. 
While such an analysis is beyond the scope of this work, we provide a conservative lower-bound estimate on $\dd\dot{N}_\varphi/\dd E_n$ by focusing on neutron interactions with materials which are common amongst modern reactor designs utilizing tritium breeding technology, as described in Sect.~\ref{subsec:reactor_summary}. 
Specifically, we focus on iron (Fe), typically present in the FW and structural supports within the wall, as well as the lithium isotopes $^{6}$Li, $^{7}$Li that are ubiquitously considered for tritium breeding in the blanket.

The total rate of spin-zero production is directly proportional to the neutron flux at the FW. 
This flux is determined by both the properties of the fusion plasma and the neutron-multiplying characteristics of the breeding blanket. 
The energy distribution and flux at the FW can vary significantly between different reactor designs and, in some cases, across neutron transport simulations. 
To account for this variability, we have considered a range of neutron spectra provided by the FISPACT-II database~\cite{fleming2018fispact}. 
For the spectra most relevant to the reactors under consideration, the general shape of the FW flux remains similar, as shown in Fig.~\ref{fig:neutron_flux}, with the largest differences appearing at low neutron energies (also known as the thermal neutron tail). 
The differences between the considered spectra result in a negligible change in the derived bounds for each production mechanism, with non-resonant production exhibiting the largest variation, changing only by a factor of two. 
Consequently, in our numerical analysis, we adopt the ITER DT spectrum as a benchmark, normalized to a power output of 2000\,MW.

Estimating $N_X$ requires explicit details on the geometry and design of the reactor walls and breeding blanket. 
Since blanket technology remains an active area of research and development, we adopt a parameterization based on generic reactor properties and keep blanket-specific parameters explicit, enabling straightforward adaptation to other reactors and blanket configurations. 
As a benchmark reactor, we consider EU DEMO~\cite{federici2014882}, whose properties and design are largely an extrapolation of the physics and technologies at ITER.  
For our estimates, we focus on the Water-Cooled Lithium-Lead~(WCLL) blanket~\cite{app112411592,martelli422752}, one of the two breeding blanket design proposals.   
Generally, blankets utilize modular breeding components to facilitate maintenance and replacement -- this motivates a parameterization of $N_X$ in terms of blanket \textit{modules} as 
\begin{equation}
    \label{eq:N_Li}
    N_X 
    = 
    \epsilon_{\rm atten.}\, \times N_{\text{module}} \times V_{\text{module}} \times \rho_{\rm blanket} \frac{f_X}{m_X} N_A\,,
\end{equation}
where $\epsilon_{\rm atten.}$ is an attenuation factor accounting for the depletion of neutrons as they traverse the reactor walls, $N_{\text{module}}$ is the total number of modules used in the reactor, $\rho_{\rm blanket}$ is the density of the blanket material where $X$ resides, $V_{\text{module}}$ is the total volume of each module, $f_X$ denotes the per-module abundance of $X$ in weight, $m_X$ is the molar mass of $X$, and $N_A$ is Avogadro's number. 
For the WCLL blanket the attenuation is estimated in Ref.~\cite{martelli422752}, and as a conservative value we take $\epsilon_{\rm atten.}=0.1$.
The full reactor consists of 48 out-board modules (and 38 in-board modules) with a plasma-facing surface area of $\approx 8.4\,$m$^2$ (poloidal height of $\approx$ 7\,m $\times$ toroidal width of $\approx$ 1.2\,m) and a thickness of $\approx 1$\,m. 
This gives 
\begin{equation}
    N_X 
    \approx 
    2.5 \times 10^{31} \,\text{cm}^3 \text{mol}^{-1} \times \rho_{\rm blanket} \frac{f_X}{m_X} \times \left(\frac{\epsilon_{\rm atten.}}{0.1}\right) \times \left(\frac{N_{\rm modules}}{48}\right) \times \left(\frac{V_{\text{module}}}{8.4 \,\text{m}^3}\right)\,.
\end{equation}
The blanket utilizes a lithium-lead~(PbLi) liquid alloy~\cite{garcinuno2022101146} as the tritium breeder, which is structurally supported with EUROFER97 steel~\cite{daum2000529}. 
For simplicity, we assume that 20\% of the module is comprised of structural material and 80\% of breeding material (by weight)~\cite{gilbert2017activation}. The breeder is composed of $\sim 84.3$\% Pb and $15.7$\% Li with 90\% $^{6}$Li enrichment, such that $f_{{^6}\text{Li}} \approx 0.8\times0.16\times0.9 = 0.11 $ and $f_{{^7}\text{Li}} \approx 0.8\times0.16\times0.1 = 0.01 $. 
The EUROFER97 steel of the supporting structure is composed of $\sim$ 89\% Fe, out of which 92\% is $^{56}$Fe, such that $f_{{^{56}}\text{Fe}} \approx 0.2\times0.89\times0.92 = 0.16$. 

We now extimate the total number of lithium and iron atoms as:
\begin{align}
    \label{eq:li6_abundance}
    N_{^6\text{Li}} 
    &\approx 
    4.8 \times 10^{30}
    \left(\frac{f_{{^6}\text{Li}}}{0.11}\right)  \left(\frac{\rho_{\rm PbLi}}{10.3 \gm\m[c]^{-3}}\right)  \left(\frac{6.02 \gm \, \text{mol}^{-1}}{m_{\rm {^6}Li}}\right)\,,   
    \\ 
    \label{eq:li7_abundance}
    N_{^7\text{Li}} 
    &\approx 
    4.6\times 10^{29} \left(\frac{f_{{^7}\text{Li}}}{0.01}\right)  \left(\frac{\rho_{\rm PbLi}}{10.3 \gm\m[c]^{-3}}\right)  \left(\frac{7.01 \gm \, \text{mol}^{-1}}{m_{\rm {^7}Li}}\right)\,, 
    \\ 
    \label{eq:Fe56_abundance}
    N_{^{56}\text{Fe}} 
    &\approx 
    5.6 \times 10^{29} \biggr(\frac{f_{{^{56}}\text{Fe}}}{0.16}\biggr)  \left(\frac{\rho_{\rm EUROFER}}{7.8 \gm\m[c]^{-3}}\right)  \left(\frac{55.9 \gm \, \text{mol}^{-1}}{m_{{^{56}}\rm Fe}}\right)\,.
\end{align}

\begin{figure}
    \centering
    \includegraphics[width=1.0\textwidth]{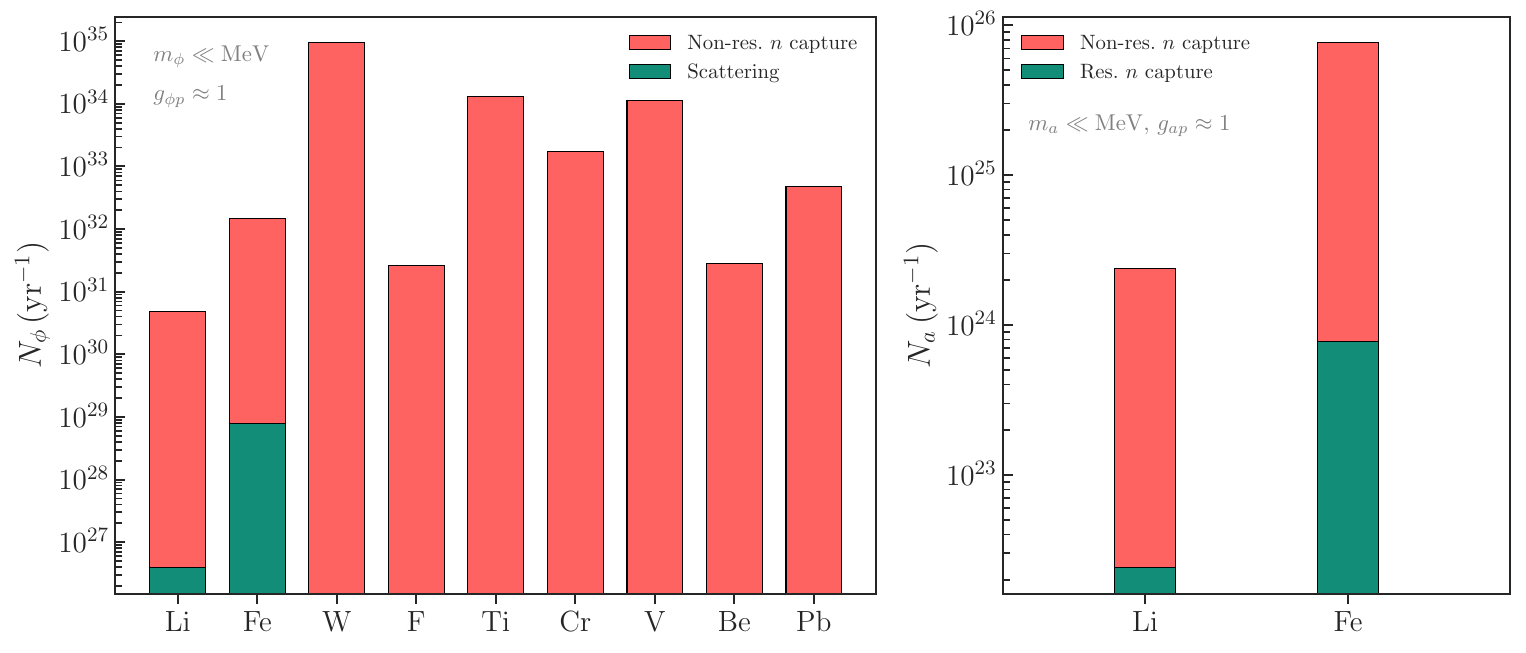}
    \caption{
    The number of light scalars (left) produced per year on $10^{29}$ nuclei of the target material (as denoted), produced either in neutron capture or neutron scattering, assuming the ITER neutron flux shape and a power output of $2000\,\text{MW}$, when setting the NP coupling $g_{\varphi p}\approx 1$. For comparison, we show ALP flux (right) for resonant and nonresonant production on Li and Fe targets under the same assumptions. 
    }
    \label{fig:NP_flux}
\end{figure}

To illustrate the sensitivity to other materials commonly present in the FW, structural supports, and blanket, we show in Fig.~\ref{fig:NP_flux} the expected number of scalar particles produced per year via neutron absorption, 
${{}^A_ZX(n,\phi){}^{A+1}_ZX}$, or via neutron scattering, ${{}^A_ZXn\to {}^{A}_ZX}n\phi$, Eqs.~\eqref{eq:scalar:n:abs}, \eqref{eq:sigm:nX}, assuming a fixed number of target nuclei. 
We use a benchmark value of $10^{29}$ nuclei of each target material, which is a ball-park estimate for the total number of nuclei in a full blanket, and assume a neutron flux energy distribution with a total power output of $2000\,\text{MW}$. 
That is, different scalar fluxes shown in Fig.~\ref{fig:NP_flux} give relative per-atom efficiencies of various targets to produce light scalars under DEMO-like conditions. 
The results shown are for massless particles, but apply in general to the case of light scalars, as long as the mass of the scalar does not approach the kinematical threshold for its production ($\sim$ few \eV[M]).
We focus on the materials relevant for the DEMO blanket ($^6$Li, $^7$Li, and $^{56}$Fe) as well as on other materials that are often considered in the context of fusion reaction. 
In the numerics, we use the natural abundances of stable isotopes. 
The results show that the total light scalar flux is relatively insensitive to the detailed composition of the blanket. 
For comparison we also show in Fig.~\ref{fig:NP_flux} (right) the flux of ALPs produced by resonant and non-resonant neutron capture, if produced on Li or Fe.

\section{Detection of light particles via deuterium association}
\label{sec:detection}

The new physics flux can be detected through a number of well-established experimental techniques involving photons ($g_{\varphi\gamma,\text{eff}}$), electrons ($g_{\varphi e}$), or nucleons ($g_{\varphi N}$). 
To maintain sole sensitivity to spin-zero couplings to nucleons, we focus on detection via deuteron dissociation sourced by $\varphi=a,\phi$, \eg,
\begin{equation}
    \varphi + d \rightarrow n + p\,,
\end{equation}
as depicted in Fig.~\ref{fig:production} (right), while in App.~\ref{app:magnetic_conv} we also give the estimates for detection via magnetic conversion. 
The dissociation process has a threshold energy of $E_{D} = 2.2 \eV[M]$, the binding energy of the deuteron. 
For the experimental setup we consider an apparatus similar to that of the Sudbury Neutrino Observatory~(SNO) experiment~\cite{SNO:2011hxd,Bhusal:2020bvx} consisting of 1000 tonnes of heavy water housed in a spherical vessel $\approx$12\,m in diameter.
The total dissociation cross section is implicitly dependent on the source, as well as on whether the spin-0 particle is a scalar or pseudo-scalar. 
We discuss the two cases in turn.

\paragraph{Pseudoscalars.} 
For an ALP-initiated dissociation, the cross section is given by Ref.~\cite{Bhusal:2020bvx}, 
\begin{align}
    \label{eq:dis_a_xsec}
     \sigma_{ad\to np} 
     = 
     \frac{g_{ap}^2}{6 m_n} \sqrt{E_a^2 - m_a^2}\frac{|\vec{k}|\alpha}{(|\vec{k}|^2 + \alpha^2)^2}\frac{(1-\alpha a_s)^2}{1+|\vec{k}|^2 a_s^2}\,,
\end{align} 
where $|\vec{k}|^2=m_n(E_a - E_d)$, $\alpha = \sqrt{m_n E_D}$, and $a_s = -23.7$\,fm is the spin-singlet scattering length. 
We also set coupling of $a$ to neutrons to zero, as in the numerical benchmarks Eqs.~\eqref{eq:benchmark:leptophobic} and~\eqref{eq:benchmark:electrons}. 

\paragraph{Scalars.} 
The cross section for scalar-induced deuteron dissociation is given by (see App.~\ref{app:scalarD} for details)
\begin{align}\label{eq:dis_phi_xsec}
    \sigma_{\phi d\to np}  
    =  
    g_{\phi p}^2  \frac{2m_n}{\sqrt{E_\phi^2 - m_\phi^2}} 
    \frac{ |\vec{k}| \alpha }{(|\vec{k}|^2+\alpha^2)^2} \frac{(1-\alpha a_s)^2}{1+|\vec{k}|^2 a_s^2}\,,
\end{align}
where $|\vec{k}|$, $\alpha$ are the same as in the ALP case and for the spin-triplet scattering length we use $a_s\approx 5.4 \m[f]$~\cite{PhysRevC.42.863}. 
As before, the couplings to neutrons are set to zero. \\[1mm]

Next, we can use the above two cross sections to arrive at the expected total number of spin-zero initiated dissociation events in the detector,
\begin{align}
    \label{eq:N_varphi_dis}
    N^{n\,\rm cap.}_{\varphi} 
    &= 
    T N_D \iint \dd E_n \dd E_\varphi \delta\left(E_\varphi -S_n-E_n\right) \frac{\dd \Phi_\varphi}{\dd E_n} ~ \sigma_{\varphi d\to np} (E_\varphi) \,, \\
    N^{2\to 2\,\rm scat.}_{\varphi} 
    &= 
    T N_D \iint \dd E_n \dd E_\varphi \delta\left(E_\varphi-E_n\right) \frac{\dd \Phi_\varphi}{\dd E_n} ~ \sigma_{\varphi d\to np} (E_\varphi) \,,
\end{align}
where $T$ is the total observation time and $N_D$ is the number of deuterium targets. 
In the case of neutron capture, for simplicity we approximate $E_n +  S_n \sim S_n$: for resonant capture this is intuitive, for non-resonant capture the region of neutron energies with the most significant contribution to radiative neutron capture, the kinetic energy is much smaller than $S_n$. This approximation has negligible impact on our results.

Using the dissociation cross sections in Eqs.~\eqref{eq:dis_a_xsec} and~\eqref{eq:dis_phi_xsec}, the spin-zero flux in Eqs.~\eqref{eq:NP_flux} and \eqref{eq:NP_num}, and taking $N_D \approx 6\times 10^{31}$, corresponding to the number of deuterium atoms in a metric ton of heavy water, we estimate the expected number of dissociation events from each production channel as:
\begin{align}
    \label{eq:N_dis_a}
    N^{\text{res.}}_a
    \approx 
    2\times 10^{20}\, g_{a p}^4 \left(\frac{T}{1\,\text{y}}\right) \left(\frac{L}{10\,\m}\right)^2,\quad
    N^{\text{non-res.}}_a
    \approx 
    4\times 10^{21}\, g_{a p}^4 \left(\frac{T}{1\,\text{y}}\right) \left(\frac{L}{10\,\m}\right)^2\,, \\
    \label{eq:N_dis_phi}
    N^{n\,\text{cap.}}_\phi 
    \approx 
    1\times 10^{31}\, g_{\phi p}^4 \left(\frac{T}{1\,\text{y}}\right) \left(\frac{L}{10\,\m}\right)^2,\quad
    N^{2\to 2}_\phi 
    \approx 
    2 \times 10^{28}\, g_{\phi p}^4 \left(\frac{T}{1\,\text{y}}\right) \left(\frac{L}{10\,\m}\right)^2\,.
\end{align}

The main SM background for spin-zero dissociation is dissociation via solar neutrinos $\nu+d\to n+p+\nu$. 
Following Ref.~\cite{Bhusal:2020bvx} using the predicted value of the ${}^8\text{B}$ solar neutrino flux in the Standard Solar Model~(SSM) and the neutral current neutrino-deuteron interaction cross section, we estimate the maximum number of background events per year due to the dissociation by solar neutrinos to be $N_\text{bkg} \sim 4800\,\text{y}^{-1}$. 
This background can, however, be studied when the reactor is not active and possibly reduced using the information about the position of the sun on the sky. 
Other backgrounds due to neutrinos from tritium and neutron decays in the reactor are negligible, since the neutrino energy is well below 2.2\,MeV which is the threshold for deuterium dissociation.

\section{Projected sensitivity to new physics}
\label{sec:results}

%
\begin{figure}
    \centering
    \includegraphics[width=0.49\textwidth]{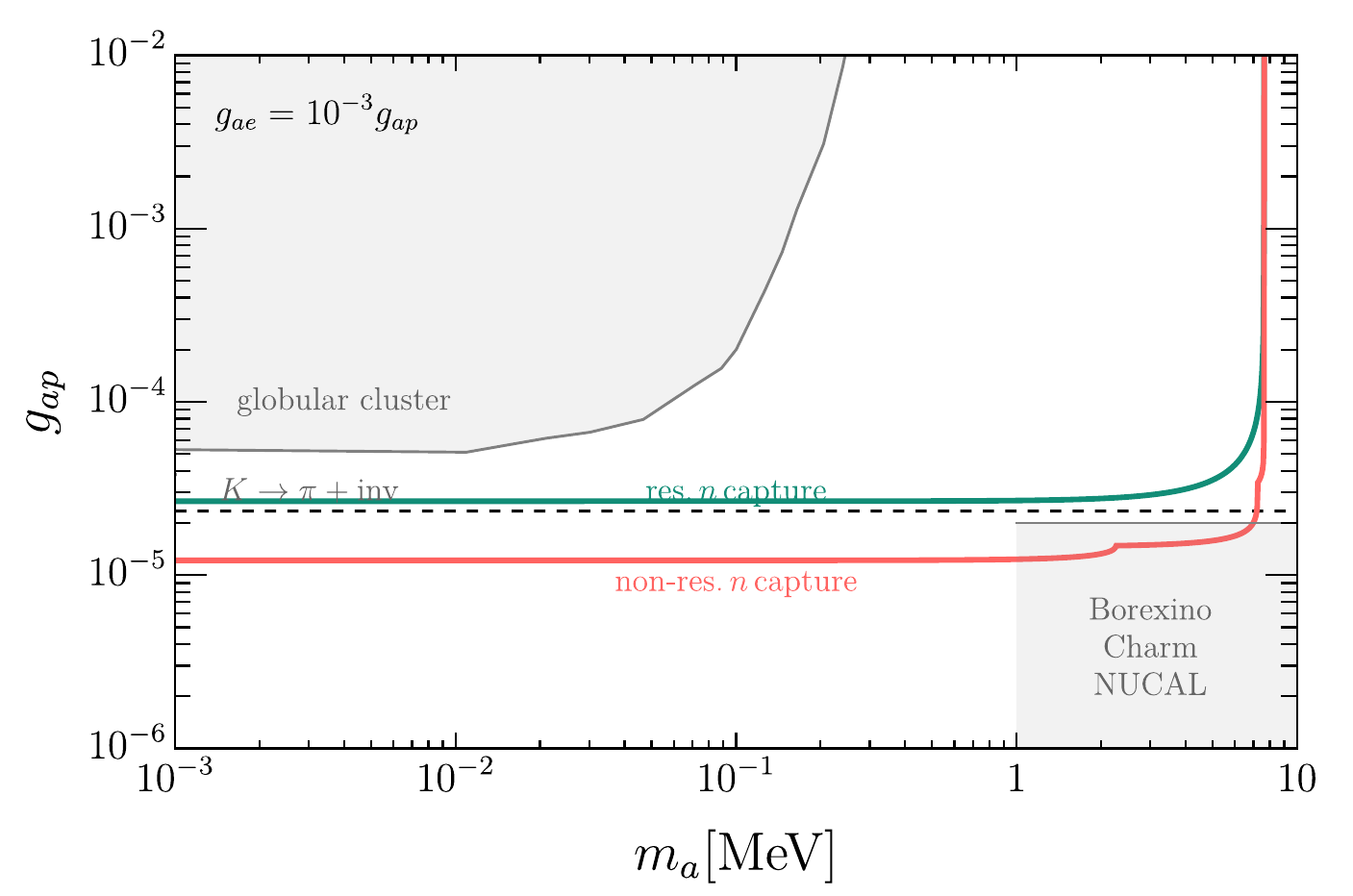}
    \includegraphics[width=0.49\textwidth]{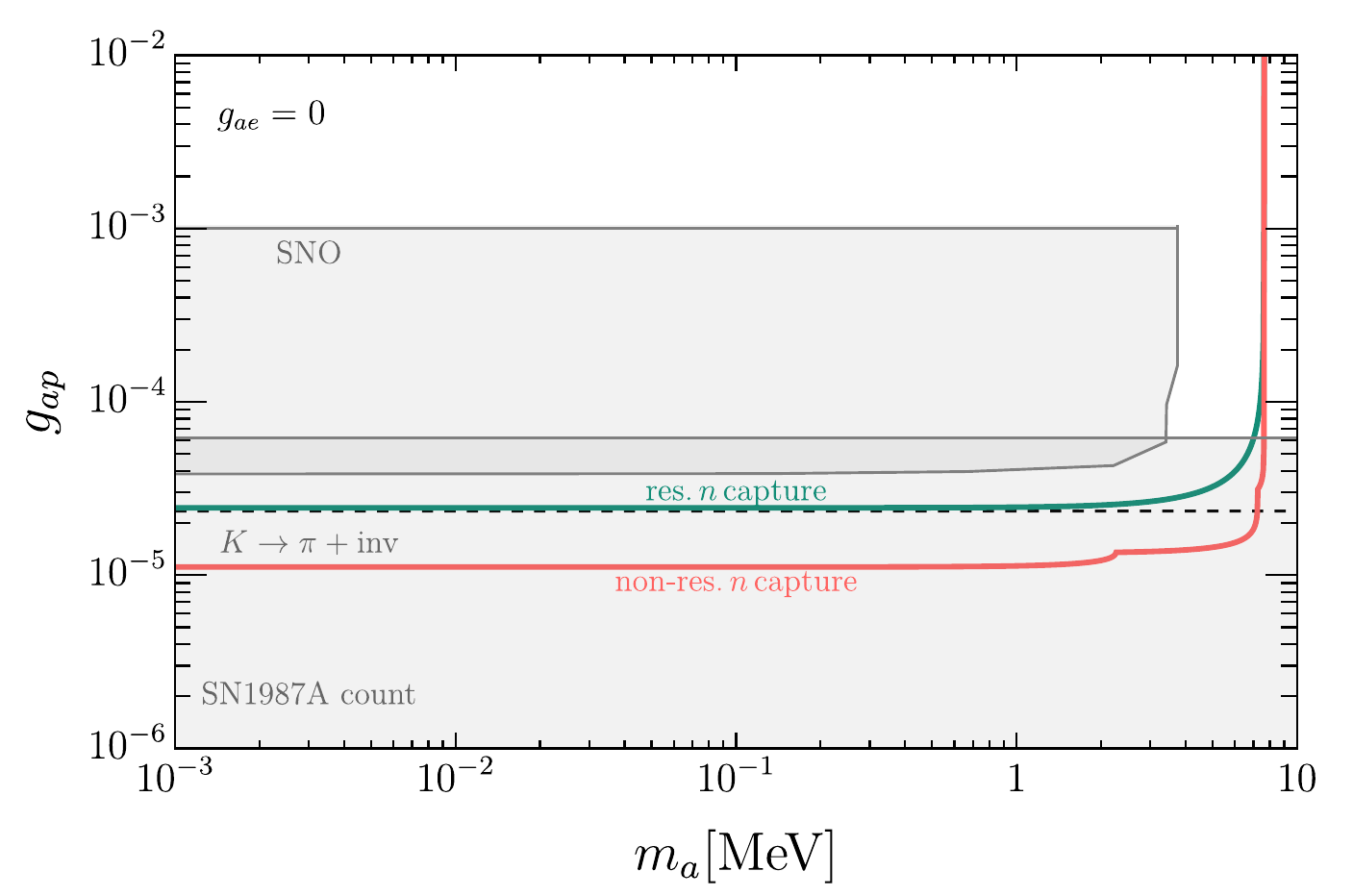}
    \caption{
    Projected sensitivities to ALP-proton coupling $g_{ap}$ from a SNO-like detector based on deuterium dissociation placed 10\,m from the center of a fusion reactor. We assume a power output of $2000\,\text{MW}$, a full blanket and the DEMO blanket composition, with proportions of $^6$Li, $^7$Li and $^{56}$Fe nuclei given in Eqs.~(\ref{eq:li6_abundance})-(\ref{eq:Fe56_abundance}). 
    Two sensitivity estimates are shown: based on ALP flux estimates due to resonant neutron capture (green solid lines), Eq.~\eqref{eq:sigma:pseudo}, and using NDA estimates for non-resonant capture rates (red solid lines), Eq.~(\ref{eq:M1:nonres}). 
    Existing constraints are shown for the two benchmarks; 
    left panel: $pe\gamma$ benchmark with ALP-electron coupling $g_{ae}= 10^{-3} g_{ap}$, Eq.~\eqref{eq:benchmark:electrons}, right panel: $p$-only benchmark with $g_{a e}=0$, Eq.~\eqref{eq:benchmark:leptophobic}. The constraints from SNO bounds on solar ALPs~\cite{Bhusal:2020bvx}, from non-observation of SN1987A induced nuclear excitations \cite{Engel:1990zd}, 
    globular cluster constraints~\cite{Ayala:2014pea,Dolan:2022kul},  Borexino~\cite{Borexino:2008wiu,Borexino:2012guz}, and the recast of CHARM~\cite{CHARM:1985anb} and NUCAL~\cite{Blumlein:1990ay}
    are denoted as shaded gray regions, while the UV model dependent bounds from  $K \rightarrow \pi  X$ searches where $X$ is invisible~\cite{NA62:2021zjw}  are denoted with dashed lines.}
    \label{fig:a_bound}
\end{figure}
\begin{figure}
    \centering
    \includegraphics[width=0.49\textwidth]{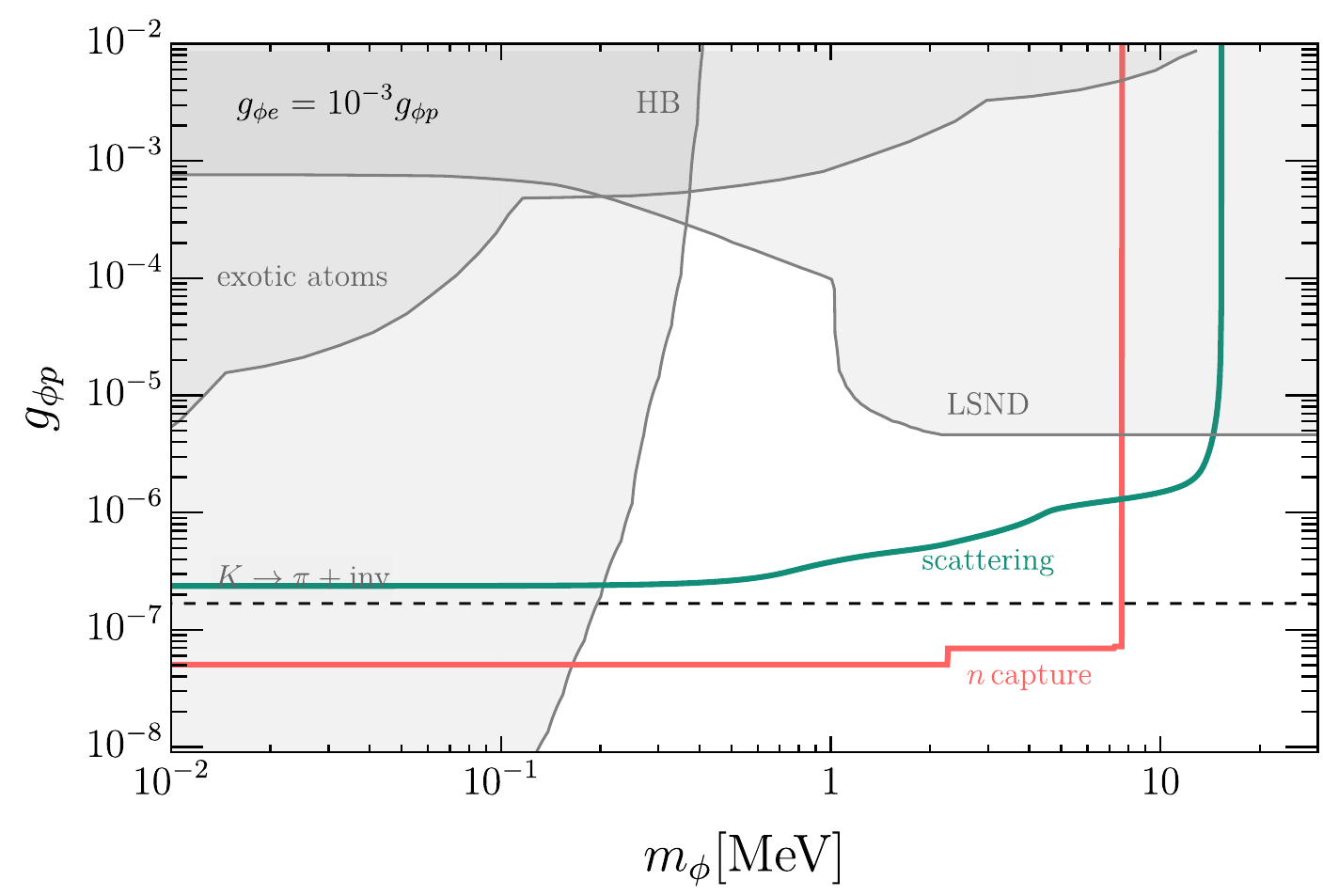}
    \includegraphics[width=0.49\textwidth]{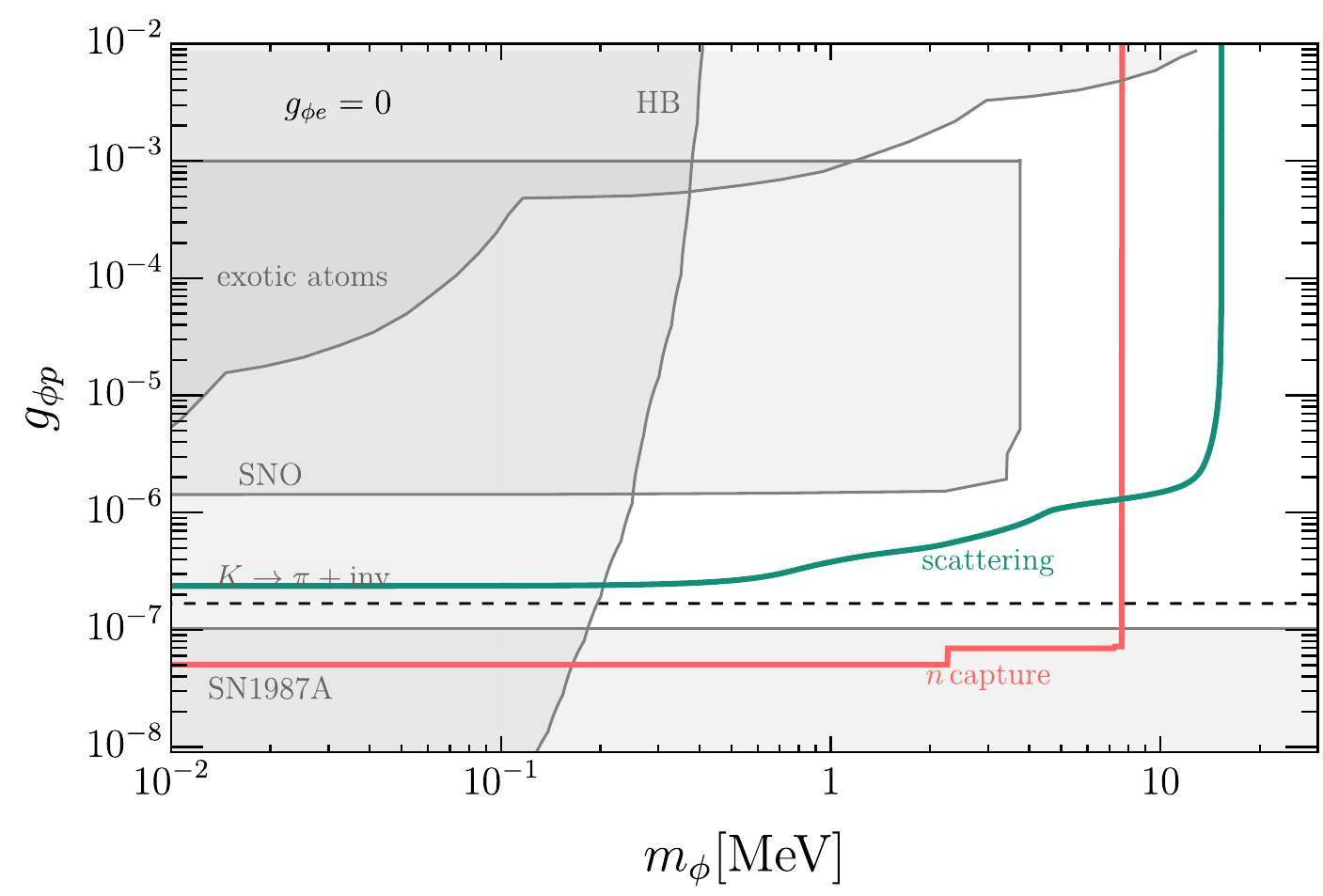}
    \caption{
   Same as in Fig.~\ref{fig:a_bound}, but for light-scalar--proton couplings, $g_{\phi p}$, where the light scalar flux estimates are based either on the NDA estimates of the 
   neutron capture rates (red line), Eq.~\eqref{eq:scalar:n:abs}, or from the $\phi$ production in $2\to 2$ scattering (green line), Eq.~\eqref{eq:sigm:nX}. Existing constraints include those shown in Fig.~\ref{fig:a_bound}, as well as the bounds from horizontal branch stars, SN1987A~\cite{Raffelt:1996wa,Benato:2018ijc}, constraints set by LSND~\cite{Pospelov:2017kep}, and constraints set using exotic atoms~\cite{ASACUSA:2016xeq,Liu:2025ows}.}
    \label{fig:phi_bound}
\end{figure}

Collecting the results from Sects.~\ref{sec:production} and~\ref{sec:detection} we can obtain the projected sensitivity to the coupling $g_{\varphi p}$ using searches at fusion reactors. 
As a benchmark we assume a SNO-like detector, using deuterium dissociation as the detection technique. 
The detector is assumed to be placed 10 meters from the center of a typical fusion facility based on deuterium-tritium fusion, with an approximate power output of $2000$\,MW and physical dimensions of the DEMO tokamak. 
The expected sensitivity after 1 year of data taking is shown in Fig.~\ref{fig:a_bound} for ALP and in Fig.~\ref{fig:phi_bound} for scalar coupling to protons. 
In the expected reach of the experiments we take into account the irreducible solar neutrino-induced background by defining the sensitivity reach as the point where $S/\sqrt{B}>2$, i.e., we require the number of signal events $N_\varphi$ to be above  $N_\varphi>2\sqrt{N_{\rm bkg}}$. 
Two different estimates for the exotic particle production rates are used both for scalars and ALPs. 

For ALP production we use either the resonant neutron absorption estimate in Eq.~\eqref{eq:sigma:pseudo} (green lines labeled ``res.~$n$ capture'') or non-resonant neutron absorption estimate in Eq.~\eqref{eq:sigm:nX}  (red lines labeled ``non-res.~$n$ capture'').
For scalar production we use the neutron absorption production estimate from Eq.~\eqref{eq:scalar:n:abs} (red lines labeled ``$n$ capture'') or neutron scattering production rate from Eq.~\eqref{eq:sigm:nX} (green lines labeled ``scattering'').
We see that searches for light particles at DEMO-like fusion facilities can probe new parameter space of such models, well beyond the currently constrained regions (grey shaded regions in Figs.~\ref{fig:a_bound} and \ref{fig:phi_bound}). 
They are also stronger than the bounds from rare kaon decays (dashed lines), for which the interpretation in terms of $g_{\varphi p}$ is model dependent (see below).

For ALPs, Fig.~\ref{fig:a_bound}, the most important constraint on $g_{ap}$ couplings in this regime was set by SNO~\cite{Bhusal:2020bvx}. 
Note that this bound does rely heavily on the assumption that the ALP-electron coupling for the solar ALPs is small enough such that they still reach the SNO detector. 
That is, the ALPs would be produced during nuclear transitions inside the Sun, and be detected in SNO via deuterium dissociation, but only if they did not decay while traversing the distance from Sun to the Earth. 
This is true for $p$-only benchmark, Eq.~\eqref{eq:benchmark:leptophobic}, in which we assume vanishing ALP-electron coupling $g_{ae}=0$. 
The resulting SNO constraint is shown in right panel in Fig.~\ref{fig:a_bound}. 
However, the SNO bound is evaded for $g_{ae}$ values such that ALPs decay before reaching SNO detector. 
As long as ALPs do not decay within reactor volume, cf. Eq.\eqref{eq:ctau:electrons}, they would still lead to a signal in the proposed search at fusion reactors, which is the case for the $pe\gamma$ benchmark with ALP-electron coupling set to $g_{ae}= 10^{-3} g_{ap}$, Eq.~\eqref{eq:benchmark:electrons} (left panel in Fig.~\ref{fig:a_bound}).
Similar to SNO constraint is the bound on hadronic axions that could be emitted in SN1987A due to their couplings to nucleons, and that would then produce observable nuclear excitations in oxygen \cite{Engel:1990zd} (right panel in Fig.~\ref{fig:a_bound}). 
The constraint however disappears, if the ALPs decay en route from SN to Earth (left panel in Fig.~\ref{fig:a_bound}).
For this ALP mass range also SN1987A cooling bounds are relevant, though they apply to smaller couplings, $g_{ap}\lesssim 10^{-6}$~\cite{Carenza:2019pxu}. 
That is, the ALP-proton coupling probed by fusion reactor experiments are large enough that the ALPs would remain trapped inside the proto-neutron star and would not contribute significantly to the SN cooling.
 
For nonzero couplings to electrons, $g_{ae}\ne0$, other constraints  become relevant as well. 
For $m_a$ in the MeV regime the most stringent is the bound from Borexino, $g_{ap} g_{ae}\gtrsim 5\cdot 10^{-13}$~\cite{Borexino:2008wiu,Borexino:2012guz} (less stringent bounds were obtained by Texono~\cite{TEXONO:2006spf}, and using BGO bolometers~\cite{Derbin:2013zba,Derbin:2014xzr}), while  for  $m_a\sim \cO(10)\,\MeV$ the recast of CHARM~\cite{CHARM:1985anb} and NUCAL~\cite{Blumlein:1990ay} experiments gives $g_{ap} g_{ae}\gtrsim 2\cdot 10^{-12}$~\cite{Waites:2022tov}. 
While for our benchmark, $g_{ae}=0$, these bounds do not apply, they should be considered if $g_{ae}$ does not vanish. 
For instance, for $g_{ae}\sim 10^{-3} g_{ap}$ they constrain $g_{ap}\gtrsim \text{few}\times 10^{-5}$. 
  
Additionally, $g_{ap}$ is also constrained from $K^+\to \pi^+ a$ searches, where $a$ is invisible and escapes detection~\cite{NA62:2021zjw}. 
This bound is UV model dependent, since it depends on the underlying origin of the ALP couplings to proton. 
In Fig.~\ref{fig:a_bound} we show the bound for the example where $g_{ap}$ is entirely due to couplings to gluons (i.e. for $c_u^a=c_d^a=c_s^a=0$ in Eq.~\eqref{eq:L:a:quarks}). 
The constraints for several other choices, such as only coupling to up quarks or down quarks can be found in~\cite{Goudzovski:2022vbt}, and are broadly comparable to the ones shown in Fig.~\ref{fig:a_bound}. 
That is, while it is possible to arrange to have cancellations between different contributions to $K^+\to \pi^+ a$ and still have a sizable $g_{ap}$, for instance by setting $c_G^a=2 c_u^a+c_d^a+c_s^a=0$ (see, e.g.,~\cite{Bauer:2021mvw}), this is cancellation is by no means generic. 
 
The relevant bounds for the light scalar benchmarks are shown in Fig.~\ref{fig:phi_bound} as grey regions. 
For the $g_{\phi p}$ coupling, we recast the SNO bound on ALP-proton coupling from Ref.~\cite{Bhusal:2020bvx}, using the solar scalar flux $\phi^{\rm sol.}_{\phi} = (g_{\phi N}^2 / 4\pi\alpha) \times 6.0\times 10^{10} \m[c]^{-2}\s^{-1}$~\cite{Pospelov:2017kep} and the scalar deuterium dissociation cross section in Eq.~\eqref{eq:dis_phi_xsec}. 
Additionally, we show constraints set using LSND~\cite{Pospelov:2017kep}, horizontal branch~(HB) stars, stellar cooling, and SN1987~\cite{Raffelt:1996wa,Hardy:2016kme,Bottaro:2023gep} (see also preliminary results in Ref.~\cite{Hardy:2024gwy}, where SN1987A cooling constraints were estimated to exclude also larger couplings).
The constraints set by the searches for $K\to \pi X$, with $X$ escaping the detector (dashed line in Fig.~\ref{fig:phi_bound}), are UV model dependent when interpreted in terms of $g_{\phi p}$, the same as was the case for ALPs. 
In Fig.~\ref{fig:phi_bound} we show the constraints for a particular case, where the couplings of $\phi$ to the SM fermions are entirely due to mixing with the SM Higgs. 
In this case the scalar-proton coupling is given by $g_{\phi p} = \sin\theta g_{hp}$, where $\sin\theta$ is the mass-mixing angle between the new scalar and the Higgs, and  $g_{hp}=-0.98(3)\times 10^{-3}$ is the effective Higgs-proton coupling (using numerical values for scalar current form factors in \eqref{eq:gphip:num}, for previous evaluations see, e.g,~\cite{Cheng:2012qr}).  
It is important to note that the $K^+\to \pi^+\phi$ bounds can deviate drastically from the bounds shown in Fig.~\ref{fig:phi_bound}, if the flavor structure of $\phi$ couplings is not the one of a Higgs-mixed light scalar. 
The general case of a light scalar coupling to quarks with flavor diagonal or almost flavor diagonal couplings was discussed recently in Ref.~\cite{Delaunay:2025lhl}. 
Using those results we see, for instance, that if the light scalar couples only to light quarks, i.e., $g_g^\phi=0$ in Eq.~\eqref{eq:Lphi:quarks}, and these couplings are proportional to quark masses, $g_u^\phi/m_u=g_d^\phi/m_d=g_s^\phi/m_s$, then the LO expressions for $K\to \pi \phi$ in chiral expansion vanish. 
In this case the constraint on $g_{\phi p}$ from $K^+\to \pi^+ \phi$ searches becomes so weak that it would not even appear in Fig.~\ref{fig:phi_bound}.

To summarize, we have seen that both for $\cO(\MeV)$ scalars and ALPs coupling to protons, there is significant discovery potential for a fusion reactor based experiment, that would probe the parameter space of dark sector models that is complementary to existing constraints. 

\section{Conclusions and outlook}
\label{sec:conclusion}

In this work, we have shown that fusion reactors will provide a promising new avenue for probing light exotic spin-0 particles, with masses up to $\cO(10\,\MeV)$. 
These particles can be produced in the mantle of fusion reactors via exotic nuclear transitions triggered by the intense neutron flux emitted from the inner volume of the reactor. 
We investigated two possible mechanisms for production of light exotic spin-0 particle: {\em  i)} neutron capture within lithium-lined breeding blankets or structural materials, and {\em ii)} bremsstrahlung of light particles in neutron scattering on nuclei. 
These processes can generate a detectable flux of exotic particles outside the reactor walls. 
Placing  a SNO-like deuterium dissociation detector near the fusion reactor, for instance, one could achieve sensitivity to new physics couplings that goes well beyond existing constraints, see Figs. \ref{fig:a_bound} and \ref{fig:phi_bound}.

The above proof-of-principle results showcase the potential of thermonuclear fusion reactors as laboratories for exploring beyond the Standard Model physics. 
However, they should also be understood for what they are: a rather preliminary exploration of the potential for fusion reactors as sources of exotic light new physics. 
First of all, several choices needed to be made in our estimates, which may or may not reflect the actual fusion reactor designs that will be eventually built -- the designs of fusion reactors are an area of active research and development, and the choices of materials will influence the production of exotic particles, see Fig.~\ref{fig:NP_flux}.
Consequently, future fusion reactor designs, particularly, those with fully operational breeding blankets and space for large-scale detectors, can be optimized to include these experimental capabilities. 
Furthermore, the theoretical predictions for the production of exotic particles can be improved, especially for the case of light scalars, where we relied on naive dimensional analysis scalings. 
Other detection techniques, beyond deuterium dissociation or magnetic conversion (discussed in App.~\ref{app:magnetic_conv}), could also be considered. 
Finally, while we focused mostly on MeV mass scale spin-0 particles, keV mass scale could also be an interesting target of exploration, both in terms of different fusion reactor based production modes as well as in term of different detection techniques, all of which could potentially lead to orders of magnitude improved sensitivities, see \cite{Dutta:2025ddv} for a recent example of an improved sensitivity using proposed detection via coherent scattering on nano-spheres.

\acknowledgments
We thank Luka Snoj and Alja\v z \v Cufar for invaluable discussions regarding ITER and for collaboration at early stages of this work. We thank Boleslaw Wyslouch, Kevin Woller for useful discussions regarding the ARC reactor.
ST was supported by the U.S. Department of Energy (DOE) Office of High Energy Physics under Grant Contract No. DE-SC0012567, and by the DOE QuantISED program through the theory consortium Intersections of QIS and Theoretical Particle Physics at Fermilab (FNAL 20-17). 
ST was additionally supported by the Swiss National Science Foundation - project n.~P500PT\_203156. 
JZ and TM acknowledge support in part by the DOE grant de-sc0011784, and NSF grants OAC-2103889, OAC-2411215, and OAC- 2417682.
The work of CB and YS, and previously that of PF, is supported by grants from the NSF-BSF (grant No. 2021800) and the ISF (grant No. 597/24).

\appendix

\section{Scalar deuterium dissociation}
\label{app:scalarD}

In this appendix we provide a calculation of light scalar induced deuterium dissociation rates, following  the treatment of $\delta$-function potentials of Ref.~\cite{Jackiw:1991je}, mirroring the ALP induced deuterium dissociation calculation in Ref.~\cite{Bhusal:2020bvx}.
 
The interaction Lagrangian $\cL=g_{\phi n} \bar{N}N\phi$, cf. \eqref{eq:Lphi:quarks}, leads to the interaction Hamiltonian 
\begin{align}
    H_I = g_{\phi n} \phi(x) \rho_N(x)\,,
\end{align}
where $\rho_N=\bar N N$ is the neutron density operator. 
To arrive at the dissociation rate we first expand the scalar field in momentum modes
\begin{align}
    \phi(x) 
    = 
    \frac{1}{\sqrt{V}} \sum_{q^\prime} \frac{1}{\sqrt{2 E_{q^\prime}}} \left(a(q^\prime) e^{i q^\prime\cdot x} + a^\dagger(q^\prime) e^{-iq^\prime\cdot x}\right)\,,
\end{align}
keeping the scaling with the volume explicit. 
The initial and final states of the $d+\phi\to p+n$ process are
\begin{align}
    \ket{i} = \ket{^3 S_1 ; q}\,, \qquad \ket{f} = \ket{^3 S^\prime_1 ; 0}\,,
\end{align}
where $^3S_1$ is the deuteron in the $^3 S_1$ ground state, $q$ is the momentum of the incoming scalar, while $^3 S^\prime_1$ is the continuum state of the $p,n$. 
The main difference between scalar dissociation and axiodissociation is that the continuum state is a spin triplet state, not a singlet. 
We now compute the interaction matrix element, which consists of the scalar part and the nucleon part. 
First for the scalar
\begin{align}
    \mel{0}{\phi(x)}{q} 
    &= \frac{1}{\sqrt{V}} \sum_{q^\prime} \frac{1}{\sqrt{2 E_{q^\prime}}} \left(\mel{0}{a(q^\prime) e^{i q^\prime\cdot x}}{q} + \mel{0}{a^\dagger(q^\prime) e^{-iq^\prime\cdot x}}{q}\right) \nonumber \\
    &= \frac{1}{\sqrt{V}} \sum_{q^\prime} \frac{1}{\sqrt{2 E_{q^\prime}}} e^{iq^\prime\cdot x} \braket{q^\prime}{q} \nonumber \\ 
    &= \frac{1}{\sqrt{V}} \frac{1}{\sqrt{2 E_{q}}} e^{iq\cdot x}\,.
\end{align}
Next, for the nucleon part, we have to evaluate $\mel{^3S_1^\prime}{\rho_N(x)}{^3S_1}$. 
For this we need to solve the Schrodinger equation for the deuteron. 
This can be reduced to a one-body problem satisfying 
\begin{align}
    \frac{1}{2m} \frac{1}{r} \pdv[2]{\,}{r} \left(r\psi(r)\right) + \left(E_D-V(r)\right)\psi(r) = 0\,,
\end{align}
Where $m = m_n/2$ is the reduced mass, $r$ is the internucleon distance, $E_D = 2.2 \eV[M]$ is the binding energy of the deuteron and $V(r) = \alpha \delta(r)$ where $\alpha = \sqrt{m_n E_D}$. The eigenfunctions are
\begin{align}
    \ket{^3S_1} = \frac{1}{\sqrt{4\pi r^2}} u(r) \chi_3 \,, \qquad
    \ket{^3S^\prime_1} = \frac{1}{\sqrt{4\pi r^2}} j(r) \chi_3\,,
\end{align}
where for the deuterium bound state $u(r)=\sqrt{2\alpha} e^{-\alpha r}$ and for the free s wave $j(r) = \sqrt{2/L} \sin(kr+\delta_0)$. 
The spin-triplet part is encoded in $\chi_3$. 
Thus, we find
\begin{equation}
\begin{split}
    \mel{^3S_1^\prime}{\rho_N(x)}{^3S_1} 
    &= \braket{\chi_3}{\chi_3} \int j^\ast(r) u(r) \dd r 
    \\
    &= \sqrt{\frac{2}{L}} \frac{\sqrt{2\alpha}\left(\alpha \sin(\delta_0) + k\cos(\delta_0) \right)}{k^2+\alpha^2} 
    \\
    &= \sqrt{\frac{2}{L}} \frac{\sqrt{2\alpha} k(1-\alpha a_s)}{(k^2+\alpha^2)\sqrt{1+k^2 a_s^2}}\,,
    \end{split}
\end{equation}
where we introduced the singlet scattering length $a_s=-1/(k \cot(\delta_0))$.
The cross section is
\begin{align}
    \sigma = \frac{2\pi V}{v_i} \left|\mel{f}{H_I}{i}\right|^2 \rho(k)\,,
\end{align}
where $\rho(k) = \dv{n}{E_k}$. 
Choosing boundary conditions $kL+\delta_0 = n\pi$, we get $\dv{n}{E_k} = \frac{L}{\pi} \dv{k}{E_k} = \frac{L}{\pi} \frac{1}{v_f}$. 
We thus find
\begin{equation}
\begin{split}
    \sigma &= g_{\phi N}^2 \frac{4}{v_i v_f} \frac{1}{E_{q}} \frac{\alpha k^2 }{(k^2+\alpha^2)^2} \frac{(1-\alpha a_s)^2}{1+k^2 a_s^2} 
     \\ 
    &=  g_{\phi N}^2  \frac{2m_n}{\sqrt{E_\phi^2 - m_\phi^2}} \frac{\alpha |\vec{k}| }{(k^2+\alpha^2)^2} \frac{(1-\alpha a_s)^2}{1+k^2 a_s^2}\,,
    \end{split}
\end{equation}
where we used $v=\frac{p}{E}$, $E_k \approx \frac{m_n}{2}$ and $q=\sqrt{E_\phi^2 - m_\phi^2}$.

The parallel expression for ALP induced deuterium dissociation, obtained in Ref.~\cite{Bhusal:2020bvx} by performing the same steps except for using the ALP derivative coupling Lagrangian and assuming the final state is $\ket{^1S_0 ; 0}$, is 
\begin{align}
    \sigma 
    = 
    \frac{1}{6} g_{aN}^2  \frac{\sqrt{E_a^2-m_a^2}}{m_n} \frac{\alpha |\vec{k}| }{(k^2+\alpha^2)^2} \frac{(1-\alpha a_s)^2}{1+k^2 a_s^2}\,.
\end{align}
This then gives for the ratio of the two cross sections
\begin{align}
    \frac{\sigma_\phi}{\sigma_a} 
    = \frac{g_{\phi N}^2}{g_{aN}^2} \frac{12m_n^2}{p_{a,\phi}^2} 
    \sim 10^5 \frac{g_{\phi N}^2}{g_{aN}^2} \left(\frac{10 \eV[M]}{p_a}\right)^2\,.
\end{align}
%

\section{Fusion reactors}
\label{app:reactors}

Here we provide a summary of forthcoming and proposed reactors that are expected to utilize tritium breeding technology, which could offer sensitivities to new physics similar to those discussed in the main text. 
For each reactor we highlight the relevant breeding blanket materials and reactor parameters relevant for spin-0 particle production in each case. 
In particular, the most important parameters for new physics production are the neutron flux (proportional to fusion power output), the choice of first wall and breeding blanket materials, and the total surface area of the inner and outer reactor walls (proportional to the total size of the reactor).  

\paragraph{ITER and DEMO.}
ITER (International Thermonuclear Experimental Reactor) is a large-scale experimental fusion reactor, currently being built in Saint Paul-les-Durance, France, that will test the scientific and technological feasibility of magnetic confinement fusion. 
ITER will serve as an important precursor to next generation fusion reactors referred to as demonstration power plants (DEMO) that will aim to produce continuous, self-sustaining fusion power and demonstrate the practical and economic viability of fusion power as a commercial energy source. 
One of the main science driver's of ITER is to evaluate tritium breeding technologies for DEMO. 
Initially, four test blanket modules (TBMs) will be tested~\cite{campbell2024} and housed along two of the reactors equatorial ports. 
The four TBMs, each nicely summarized by their name, are the European Water-Cooled Lithium-Lead~(WCLL)~\cite{aubert2020design}, the European Helium-Cooled Pebble-Ceramic~(HCCP)~\cite{zmitko2018development}, the Japanese Water-Cooled Ceramic-Breeder~(WCCB)~\cite{tanigawa2018cylindrical}, and the Chinese Helium-Cooled Ceramic Breeder~(HCCB)~\cite{wang2019current}.
After the initial testing run for the above mentioned breeding modules, additional breeding blankets may be tested in the future~\cite{swami2022neutronic}. 
When constructed, ITER will be the largest fusion reactor ever built with major and minor radii of 6.2~m and 2~m, respectively. The total fusion output is expected to range between 350-400\,MW depending on the running mode. 
The next generation DEMO reactors are expected to be significantly larger, with major and minor radii between 8-9~m and 2-3~m, respectively. 
With fully operational blankets these reactors are projected to produce fusion outputs between 1500 to 2500~MW. 

\paragraph{CFETR.} The Chinese Fusion Engineering Testing Reactor~(CFETR) is a proposed tokamak designed to bridge the gap between ITER and future fusion-based power plants. 
CFETR aims to validate key technologies required for sustained fusion energy production, including tritium self-sufficiency, advanced plasma control, and high-efficiency power generation~\cite{liu2019progress,liu2019updated,liu2022design}. The proposed design for the reactor will utilize 2\,mm thick, castellated tungsten armor tiles as the FW. 
Beyond the FW, 10\,mm thick U-shaped plates made of Oxide Dispersion-Strengthened~(ODS) steel will house the tritium breeding blanket material. 
The blanket utilizes a mixed pebble bed of lithium orthotitanate (Li$_2$TiO$_3$) and beryllium inter-metallic compound (Be$_{12}$Ti). 
The total number of modules in the WCCB blanket system is 432, divided into 16 toroidal blanket sectors, each comprising multiple inboard and outboard modules. 
The current concept design sets the reactor size to have major and minor radii of 7.2\,m and 2.2\,m, respectively, with an expected final phase power output of $\approx 1000$\,MW. 

\paragraph{ARC.}
The Affordable Robust Compact~(ARC) reactor aims to deliver power comparable to ITER/DEMO at a significantly reduced size and cost~\cite{sorbom2015arc,kuang2018conceptual}. 
The conceptual reactor design places the major and minor radii at 3.3\,m and 1.13\,m, respectively, with a total fusion output of roughly 500\,MW.
The design proposal utilizes a first wall composed of 10\,mm thick tungsten layer followed by 1\,cm of structural material consisting of a vanadium alloy VCr$_4$Ti${_4}$ (many other materials have been considered, see Refs.~\cite{bocci2020arc,pettinari2022arc}).
Behind this structure is a 3\,cm blanket cooling channel aiding in heat extraction and supporting the thermal regulation of the system. 
This is followed by another 3\,cm layer of structure VCr$_4$Ti${_4}$. 
The final bulk region of the blanket, measuring approximately 100\,cm in thickness, consists of the FLiBe molten salt that acts as a tritium breeder, neutron moderator, and a shield for the high-temperature superconducting magnets. 
Note that many other materials have also been investigated for the blanket materials, see Ref.~\cite{segantin2020optimization} for more details. 
In total, the overall thickness of the reactor components amounts to 107.1\,cm, including all structural and functional layers.

\section{Magnetic conversion}
\label{app:magnetic_conv}

\begin{figure}[t!]
    \centering
    \includegraphics[width=0.49\linewidth]{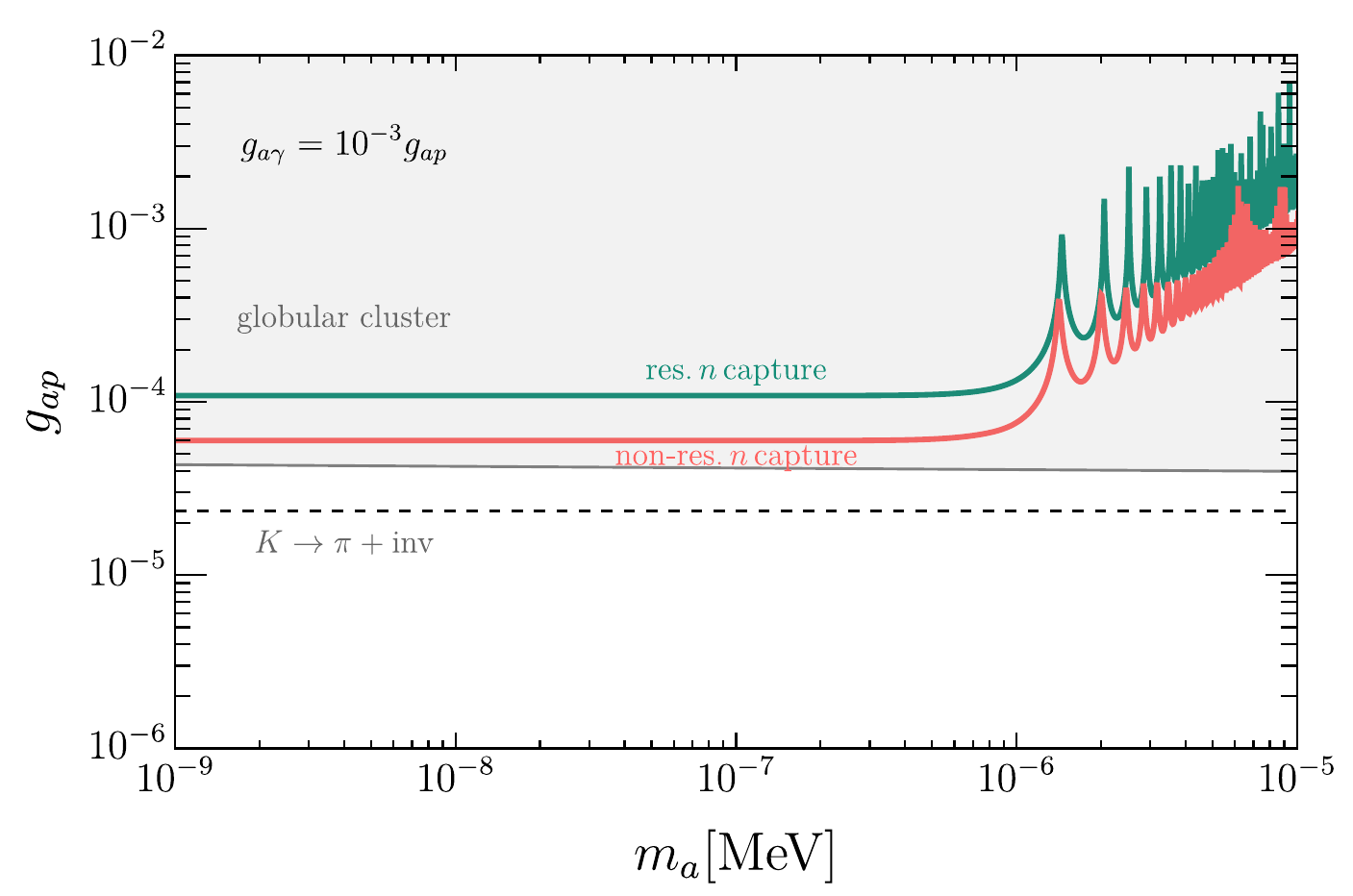}
    \includegraphics[width=0.49\linewidth]{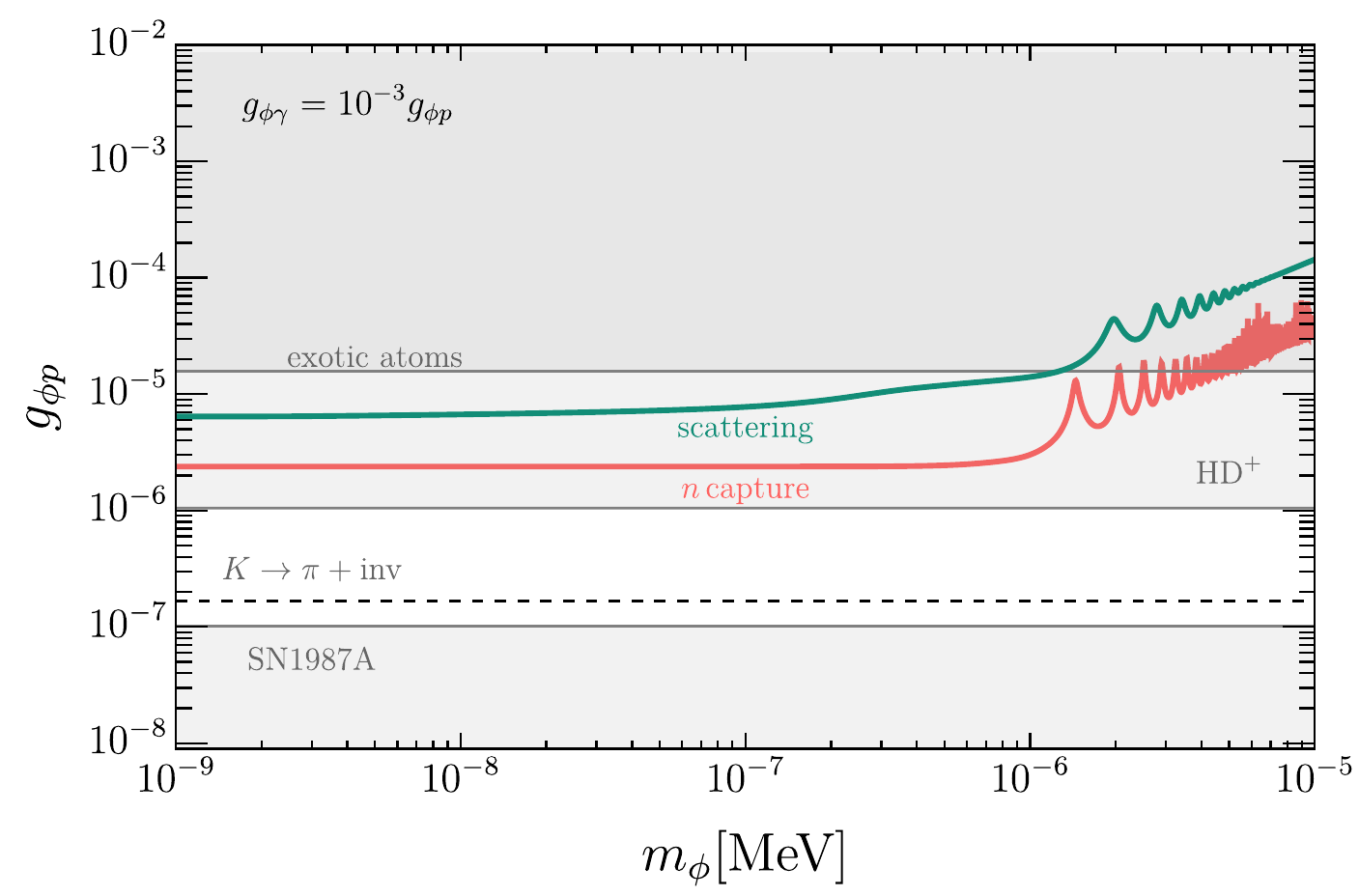}
    \caption{The expected sensitivity to ALP couplings to protons $g_{ap}$ (left) and  scalar coupling to protons $g_{\phi p}$ (right), for the $pe\gamma$ benchmark, Eq.~\eqref{eq:benchmark:electrons}, using magnetic conversion in a CAST-like detector. For ALPs we indicate the existing bounds from SNO~\cite{Bhusal:2020bvx} (grey region) and the UV dependent bounds from  $K \rightarrow \pi  X$ searches where $X$ is invisible~\cite{NA62:2021zjw} (dashed line). For scalars the UV dependent existing bounds from $K \rightarrow \pi X$ where the scalar couples to the SM via Higgs mixing are shown (dashed line), as well as the SN1987 constraints and the globular cluster constraints~\cite{Raffelt:1996wa,Benato:2018ijc} (grey regions). }
    \label{fig:CAST_bounds}
\end{figure}

The CERN Axion Solar Telescope (CAST) experiment~\cite{CAST:2017uph} aims to detect solar ALPs by converting them to photons in a magnetic field and then detecting these gamma rays in a calorimeter, which is engineered to detect $\gamma$-rays across a broad spectrum from tens of keV to several MeV.
As in Eq.~\eqref{eq:N_varphi_dis}, the number of photons  produced at the CAST-like detector by $\varphi$ conversions, which are produced either through neutron capture or elastic scattering, $N_{\varphi,\gamma}$, can be estimated as~\cite{Arias-Aragon:2023ehh}
\begin{align}
    N^{n\, \rm scat.}_{\varphi,\gamma} 
    &= 
    t_{\rm obs} ~ S ~ \epsilon 
    \iint \dd E_n \dd E_\varphi \delta\left(E_\varphi -S_n-E_n\right) \frac{\dd \Phi_\varphi}{\dd E_n} ~  \cP_{\varphi \to \gamma}(E_\varphi) \,, \\
    N^{2\to 2\,\rm scat.}_{\varphi,\gamma} 
    &= 
    t_{\rm obs} ~ S ~ \epsilon \iint \dd E_n \dd E_\varphi \delta\left(E_\varphi-E_n\right) \frac{\dd \Phi_\varphi}{\dd E_n} ~ \cP_{\varphi \to \gamma}(E_\varphi) \,,
\end{align}
where $t_{\rm obs}$ is the observation time, $S$ the cross-sectional area of the detector, $\epsilon$ the detector efficiency and $\cP_{\varphi \to \gamma}$ the conversion probability of an ALP/scalar in a magnetic field of length $L$ and strength $B$, which is given by~\cite{vanBibber:1988ge}
\begin{align}
   \cP_{\varphi \to \gamma} 
   = 
   \left(\frac{g_{\varphi\gamma} B}{q}\right)^2\sin^2\left(\frac{qL}{2}\right),
\end{align}
with $q=\sqrt{(m_\varphi^2/(2E_\varphi))^2+(g_{\varphi\gamma}B)^2}$. 
Note that the detection efficiencies of scalar and pseudoscalar particles in helioscopes are exactly the same, \eg~\cite{Flambaum:2022zuq}, given that the Primakoff production cross sections are identical for the same strengths of couplings to photons in the two cases. 

For detection via magnetic conversion, we consider a detector with similar specifications as CAST~\cite{CAST:2017uph}, i.e., we use $B=9\,\rm T$, $L=9\,\rm m$, $S=14\,\rm cm^2$, $\epsilon=56 \%$,\footnote{We note that the detector efficiency calculated in Ref.~\cite{CAST:2009klq}, is calibrated with respect to the $^7\rm{Li}^* \to ^7\rm{Li} + \gamma$ $M_1$ transition which occurs for $E_\gamma^{\rm M1} = 478\,\keV$. We have assumed similar efficiencies for the E1 channel of $^6\rm{Li}$ as well as the M1 channel of $^8\rm{Li}$. } 
and similar background estimates, amounting to 1 event per second~\cite{CAST:2009klq}. 
Allowing for 1 year of observation time, we derive bounds on the parameter space of the new physics models by requiring $N_\gamma > \sqrt{N_{\rm bkg}} = \sqrt{t_{\rm obs}~(\rm{1/s})} \sim 10^4$. 
The results for the $pe\gamma$ benchmark, Eq.~\eqref{eq:benchmark:electrons}, are shown in Fig.~\ref{fig:CAST_bounds} for ALPs (left) and scalars (right). 
While the projected reach of such an experimental set-up is weaker then the present constraints for this particular benchmark, a more thorough analysis of full parameter space is warranted, see also discussion in Sect.~\ref{sec:results} in the main text. 

\section{Naive estimates for light scalar production - a toy example}
\label{sec:toy:example}

To obtain intuition regarding rescaling of the non-resonant emission rates for photons to the rates of emissions of light spin-$0$ particles $\varphi=a, \phi$ we consider a toy example, an absorption of a light electrically neutral spin-$0$ nucleus $s_n$, on a heavier spin-1/2 nucleus ${\cal N}$, creating a heavy spin 1/2 nucleus ${\cal N}'$, where both nuclei have charge $Z=+1$. 
The choice of spins is arbitrary, and was made to simplify the presentation of results. 
The light nucleus $s_n$ is a stand-in for a neutron in the processes we are interested in the main text, and thus we assume also for the toy model that there is a large hierarchy between the masses, $m_{s_n}\ll m_{\cal N}, m_{{\cal N}'}$.

We consider three types of $s_n$ absorption rates:
\begin{itemize}
\item ${\cal N} (s_n, \gamma) {\cal N}'$: This transition proceeds through SM interactions. 
At the level of nuclei, we use the following effective interaction Lagrangian
\beq
    \cL_{\rm int}
    = 
    (g_n \bar {\cal N}' {\cal N} s_n +{\rm h.c.}) +i e \bar {\cal N} \slashed A{\cal N}+i e \bar {\cal N}' \slashed A{\cal N}',
\eeq
where the interaction with the photon comes from the kinetic Lagrangian for fields ${\cal N}$, ${\cal N}'$.

\item ${\cal N} (s_n, \phi) {\cal N}'$: The ${\cal N}'{\cal N} s_n$ interaction is the same as before, while the exotic transitions ${\cal N}'{}^*\to {\cal N}' \phi$, $s_n^* \to s_n \phi$ (where ${\cal N}'{}^*$ and $s_n^*$ are off-shell states) proceed through new scalar currents, giving 
\beq
    \cL_{\rm int}
    = 
    (g_n \bar {\cal N}' {\cal N} s_n +{\rm h.c.}) +g_\phi \bar {\cal N}' {\cal N}' \phi +g_\phi m_{s_n}  s_n^2 \phi.
\eeq
For simplicity, the light exotic particle $\phi$ was assumed to couple to ${ s}_n$ and ${\cal N}'$, but not to ${\cal N}$ (our conclusions do not change qualitatively, if this is relaxed). 
Note that the properly normalized exotic scalar currents have the same strength for emissions of $\phi$ from ${\cal N}'$ and from $s_n$, which is the appropriate long wavelength limit in the case that ${\cal N}$ does not couple to $\phi$. 

\item ${\cal N} (s_n, a) {\cal N}'$: The effective Lagrangian is in this case assumed to be
\beq
    \cL_{\rm int}
    = 
    (g_n \bar {\cal N}' {\cal N} s_n +{\rm h.c.})+g_a \big(\bar {\cal N} i \gamma_5 {\cal N}\big) a+g_a' \big(\bar {\cal N}' i \gamma_5 {\cal N}'\big) a,
\eeq
where we take $g_a\ne g_a'$, since these couplings depend on the internal structure of nuclei. 
\end{itemize}

We are interested in the non-relativistic limit. 
The heavy particle effective theory expressions for the above interactions terms follow from the following non-relativistic reductions  (see, e.g., Refs.~\cite{Manohar:2000dt,Bishara:2017pfq})
\begin{align}
    \bar {\cal N}' {\cal N} s_n & 
    \to \bar {\cal N}_v' {\cal N}_v s_n +\cO(1/m^2),\\
    \label{eq:sn:scalar}
    m_{s_n} s_n^2 \phi  & 
    \to s_{n,v}^\dagger s_{n,v} \phi  +\cO(1/m_{s_n}^2),\\
    \label{eq:N:scalar}
    \bar {\cal N} {\cal N} \phi  & 
    \to \bar {\cal N}_v {\cal N}_v \phi +\cO(1/m_{\cal N}^2),\\
    \bar {\cal N} i \gamma_5 {\cal N}  a &
    \to -\frac{i}{m_{\cal N}} \bar {\cal N}_v q\cdot S_{\cal N} {\cal N}_v a +\cO(1/m_{\cal }^2)\\
    \begin{split}
    \label{eq:em: interactions}
    A_\mu \bar {\cal N} \gamma^\mu {\cal N}  &\to \bar {\cal N}_v {\cal N}_v v\cdot A+\bar {\cal N}_v {\cal N}_v \Big( \frac{p_{1\perp}^\mu+p_{2\perp}^\mu}{2m_{\cal N}}\Big) A_\mu \\
    &\quad+\frac{i}{m_{\cal N}} \epsilon^{\mu\nu\alpha\beta} v_\alpha q_\nu \big(\bar {\cal N}_v S_{N\beta} {\cal N}_v\big) A_\mu +{\mathcal O}(1/m_{\cal N}^2).
    \end{split}
\end{align}
The same expressions as in Eqs.~\eqref{eq:N:scalar}-\eqref{eq:em: interactions} apply also for ${\cal N'}$, with the replacements ${\cal N}\to {\cal N}'$. 
Above, ${\cal N}_v$ and $s_{n,v}$ are the heavy particle (i.e., non-relativistic) fields,  $S^\mu=\gamma_\perp^\mu \gamma_5/2$ is the spin operator, $v^\mu=(1,\vec 0)$ is the four-velocity of the lab frame, $p_{1(2)\perp}$ are the soft momenta of incoming (outgoing) ${\cal N}$ particle so that full momentum is $m_{\cal N} v^\mu +p_{\cal N}^\mu$,  $p_{{\cal N}\perp}^\mu=p_{{\cal N}}^\mu - (v\cdot p_{{\cal N}})v^\mu$ and similarly for ${\cal N}'$ and $s_n$, while $q=p_2-p_1$ is the four-momentum exchange. 
The masses for the heavy particles satisfy $m_{s_n}+m_{{\cal N}}= m_{{\cal N}'}+S_n$, where $S_n$ is the energy release in the limit of vanishing velocities (in the main text this becomes the neutron separation energy). 

Note that the $\bar {\cal N}_v {\cal N}_v (v\cdot A)$ term in Eq.~\eqref{eq:em: interactions} does not contribute to the production of on-shell photons.  Furthermore, the $1/m_{\cal N}$ suppressions were obtained under the assumption that ${\cal N}$ is an elementary particle. 
For composite objects, such as nuclei, this is not necessarily true. 
For instance, for the terms that are not fixed by re-parametrization invariance, the correct size is given by the typical spacial extent of the nucleus. 
If, instead, the above toy model is replaced with couplings to single nucleon currents, the first interaction term in Eq.~\eqref{eq:em: interactions} would be due to $A_\mu$ coupling to convective current, and the second due to intrinsic magnetization in the nucleus.

For the ratios of the cross sections one gets
\beq
    \frac{(d\sigma)_\varphi}{(d\sigma)_{\rm E1}}
    =
    \frac{(\sum M M^*)_\varphi}{(\sum M M^*)_{\rm E1}}\frac{(d\text{P.S.})_\varphi}{(d\text{P.S.})_\gamma},
\eeq
where we only used the E1 part of the electromagnetic cross section.
The ratio of phase space factors is
\beq
    \frac{(d\text{P.S.})_\varphi}{(d\text{P.S.})_\gamma}=\frac{|\vec{p}_\varphi|}{|\vec{p}_\gamma|}= \frac{(E_\varphi^2-m_\varphi^2)^{1/2}}{E_\gamma}.
\eeq
Taking the limit $m_{\cal N}, m_{{\cal N}'}\gg m_{s_n}\gg m_\phi$ and keeping only the leading terms in $1/m_{{\cal N},{\cal N}', s_n}$ expansion,  one obtains for the scalar emission 
\beq
    \begin{split}
    \label{eq:scaling:phi:E1}
    \frac{(d\sigma)_\phi}{(d\sigma)_{\rm E1}}
    &=
    \left(\frac{g_\phi}{e}\right)^2 
    \left(\frac{m_{{\cal N}}}{|\vec p_{\cal N}|}\right)^2 
    \left(\frac{E_{s_n}}{ E_\phi -E_{s_n}}\right)^2 
    \frac{\sqrt{E_\phi^2 -m_\phi^2}}{E_\phi}
    \end{split}
\eeq
where we set $E_\phi=E_\gamma$, with $E_{s_n}$ the kinetic energy of $s_n$ in the initial state,  scattering on ${\cal N}$ at rest. The first factor, $\big(g_\phi/{e}\big)^2$ is simply the ratio of coupling strengths for $\phi$ vs.~$\gamma$ emission. 
The $\big({m_{{\cal N}}}/{|\vec p_{\cal N}|}\big)^2$ factor encodes the additional suppression of E1 electromagnetic emission from external legs, compared to the emission of a scalar particle $\phi$. In this estimate it was crucial that we assumed that nucleons are point particles. 

If the electromagnetic E1 transition is due to the decay of an intermediate nuclear resonance to a ground state, as is the case in the main text, the transition is due to a collective effect of emissions from protons inside the nucleus arranging into a new ground state. 
The $\big({|\vec p_{\cal N}|}/{m_{{\cal N}}}\big)^2$ suppression of E1 transition then becomes instead $ \sim \big({|\vec q|}/m_p\big)^2$. 
The appearance of proton mass $m_p$ signifies that the transitions are dominated by single proton currents, while $|\vec q|=E_\gamma$ is the momentum of the outgoing photon. 
The appearance of net momentum flow out of the nuclear system, instead of a typical individual proton momentum inside a nucleus, is a result of the fact that all momenta of constituents need to sum up to a nucleus at rest. 
In the NDA estimates in the main text, Eq.~\eqref{eq:scalar:n:abs}, we conservatively used a slightly less effective suppression of E1 transitions, replacing $ \sim \big(E_\gamma/m_p\big)^2$ with $ \sim \big(E_\gamma/Q \big)^2$, where $Q\sim 250\,\MeV$ is the inverse of a typical nuclear radius, following the NDA estimates in~\cite{Donnelly:1978tz}. 

Note that had we used different assumptions in the toy model, for instance that $\phi$ only couples to ${\cal N}$, ${\cal N}'$ and not at all to $s_n$, our main conclusions would not change. 
In this case one would obtain for the ratio of cross sections the result in Eq.~\eqref{eq:scaling:phi:E1}, but with the replacement $E_{s_n}\to m_{s_n} E_{s_n}/m_{\cal N}$. 
However, once one identifies again that the relevant mass suppression scale in the nuclear transition of interest in the main text is $m_p$, and thus taking the replacement limits $m_{\cal N}, m_{s_n}\to m_p$ and $|\vec p_{\cal N}|\to |\vec q|$, this new result still leads to the same NDA parametric scalings in Eq.~\eqref{eq:scalar:n:abs} as before. 

For the pseudoscalar case we have
\beq
    \frac{(d\sigma)_a}{(d\sigma)_{\rm E1}}
    =
    \Big(\frac{g_a-g_a'}{e}\Big)^2 \biggr(\frac{|\vec p_a|}{|\vec p_{\cal N}|}\biggr)^2 \frac{(E_a^2-m_a^2)^{1/2}}{E_a},
\eeq
where, similarly as the scalar, we set $E_a=E_\gamma$. 
If one is interested in a resonant neutron absorption, the suppression of E1 transition is, as argued above, given by $|\vec q|=E_\gamma$ and thus $|\vec p_{\cal N}|$ is to be replaced by $E_\gamma=E_a$, giving the same parametric scaling as the exact result in Eq.~\eqref{eq:M1_flux} in the main text (in the main text the normalization is to the M1 electromagnetic transition, which does not change the parametrics).

Using the toy model we can also estimate the probability for a soft emission of a light scalar in the neutron scattering. 
That is, one can relate the $2\to 3$ scattering $s_n {\cal N}' \to s_n {\cal N}' \phi$, which involves an emission of $\phi$, to a  $2\to 2$ scattering process, $s_n {\cal N}' \to s_n {\cal N}'$. 
For the amplitudes we have
\beq
    M_{2\to 3}
    \simeq 
    -i M_{2\to 2} g_{\phi} \frac{E_{s_n}}{E_\phi (E_{s_n}-E_\phi)},
\eeq
where $E_{s_n}$ is the kinetic energy of the incoming light nucleus $s_n$, with ${\cal N}$ at rest. 
Above, we used that in most of phase space $ E_\phi \sim\cO\left(E_{s_n}\right)$, while the outgoing $s_n$ and ${\cal N}$ share the momentum recoil, and thus the $s_n$ in the final state has a typical kinetic energy that is $\cO(E_{s_n}^2/m_{s_n})\ll E_\phi$. 
Note also, that the leading contributions comes from the intermediate off-shell ${\cal N}$ state. 
Including the phase space factor gives
\beq
    \label{eq:dsigma23:app}
    d \sigma_{2\to 3}
    \simeq 
    d\sigma_{2\to 2} \biggr(\frac{E_{s_n}}{E_{s_n}-E_\phi}\biggr)^2 \frac{g_\phi^2}{4 \pi^2}\frac{p_\phi^2 dp_\phi}{E_\phi^3},
\eeq
where $ d\sigma_{2\to 2} $ is the differential cross section for $2\to 2$ scattering, which includes the remaining phase space differential factors for the two heavy particles in the final state. 
The scaling in Eq.~\eqref{eq:dsigma23:app} agrees parametrically with the NDA estimate in Eq.~\eqref{eq:sigm:nX} in the main text, where the coupling of $\phi$ to nucleus is given by $Z g_{p\phi}$ in the long wavelength limit. 

\bibliographystyle{JHEP}
\bibliography{fusion_ALPs}

\end{document}